\documentclass[english,aps,pre,preprint,superscriptaddress,nofootinbib]{revtex4}
\usepackage{amsmath}
\usepackage{graphicx}
\usepackage{amssymb}
\usepackage{ae,aecompl}
\usepackage{babel}

\providecommand{\onlinecite}{\cite}

\begin{document}

\title{Pair Correlation Function Characteristics of Nearly Jammed Disordered
and Ordered Hard-Sphere Packings}

\author{Aleksandar Donev}

\affiliation{\emph{Program in Applied and Computational Mathematics}, \emph{Princeton
University}, Princeton NJ 08544}

\affiliation{\emph{Materials Institute}, \emph{Princeton University}, Princeton
NJ 08544}

\author{Salvatore Torquato}

\email{torquato@electron.princeton.edu}

\affiliation{\emph{Program in Applied and Computational Mathematics}, \emph{Princeton
University}, Princeton NJ 08544}

\affiliation{\emph{Materials Institute}, \emph{Princeton University}, Princeton
NJ 08544}

\affiliation{\emph{Department of Chemistry}, \emph{Princeton University}, Princeton
NJ 08544}

\author{Frank H. Stillinger}

\affiliation{\emph{Department of Chemistry}, \emph{Princeton University}, Princeton
NJ 08544}

\begin{abstract}
We study the approach to jamming in hard-sphere packings, and, in
particular, the pair correlation function $g_{2}(r)$ around contact,
both theoretically and computationally. Our computational data unambiguously
separates the narrowing delta-function contribution to $g_{2}$ due
to emerging interparticle contacts from the background contribution
due to near contacts. The data also shows with unprecedented accuracy
that disordered hard-sphere packings are strictly isostatic, i.e.,
the number of exact contacts in the jamming limit is exactly equal
to the number of degrees of freedom, once rattlers are removed. For
such isostatic packings, we derive a theoretical connection between
the probability distribution of interparticle forces $P_{f}(f)$,
which we measure computationally, and the contact contribution to
$g_{2}$. We verify this relation for computationally-generated isostatic
packings that are representative of the maximally jammed random state.
We clearly observe a maximum in $P_{f}$ and a nonzero probability
of zero force, shedding light on long-standing questions in the granular-media
literature. We computationally observe an unusual power-law divergence
in the near-contact contribution to $g_{2}$, persistent even in the
jamming limit, with exponent $-0.4$ clearly distinguishable from
previously proposed inverse square root divergence. Additionally,
we present high-quality numerical data on the two discontinuities
in the split-second peak of $g_{2}$, and use a shared-neighbor analysis
of the graph representing the contact network to study the local particle
clusters responsible for the peculiar features. Finally, we present
the first computational data on the contact-contribution to $g_{2}$
for vacancy-diluted FCC crystal packings and also investigate partially
crystallized packings along the transition from maximally disordered
to fully ordered packings. Unlike previous studies, we find that ordering
has a significant impact on the shape of $P_{f}$ for small forces.
\end{abstract}
\maketitle

\newcommand{\Cross}[1]{\left|\mathbf{#1}\right|_{\times}}

\newcommand{\CrossL}[1]{\left|\mathbf{#1}\right|_{\times}^{L}}

\newcommand{\CrossR}[1]{\left|\mathbf{#1}\right|_{\times}^{R}}

\newcommand{\CrossS}[1]{\left|\mathbf{#1}\right|_{\boxtimes}}

\newcommand{\V}[1]{\mathbf{#1}}

\newcommand{\M}[1]{\mathbf{#1}}

\newcommand{\D}[1]{\Delta#1}

\newcommand{\sV}[1]{\boldsymbol{#1}}

\newcommand{\sM}[1]{\boldsymbol{#1}}

\newcommand{\grad}{\boldsymbol{\nabla}}

\newcommand{\ReducedItemSeparation}{\setlength{\itemsep}{0ex}}

\section{Introduction}

Jamming in hard-sphere packings has been studied intensely in the
past years. In this paper, we investigate the pair correlation function
$g_{2}(r)$ of the classical three-dimensional hard-sphere system
near a jamming point for both disordered (amorphous, often called
random) as well as ordered (crystal) jammed packings. The basic approach
follows that of Ref. \cite{FreeVolume_ClosePacked}, developed further
for crystal packings of rods, disks and spheres in Ref. \cite{LimitingPolytope_HS}.
We focus on \emph{finite} sphere packings that are almost \emph{collectively
jammed} \cite{Torquato_jammed,Jamming_LP}, in the sense that the
configuration point is trapped in a very small region of configuration
space around the point representing the jammed \emph{ideal} packing
\cite{Jamming_LP}. Difficulties with extending the results to infinite
packings will be discussed in what follows. In the ideal jammed packing
particle contacts necessary to ensure jamming are exact, and the particles
cannot at all displace, even via collective motions. Such ideal jammed
(or rigid) packings have long been the subject of mathematical inquiry
\cite{Connelly_disks}; however, they are not really attainable in
numerical simulations where produced packings invariably have some
interparticle gaps (even taking into account the unavoidable roundoff
errors). It is therefore instructive to better understand the approach
to this ideal jammed state computationally and theoretically, which
is the primary objective of this paper.

We choose as our main tool of exploration the shape of the venerable
orientationally-averaged pair correlation function $g_{2}(r)$ around
contact. This is because this function is a simple yet powerful encoding
of the distribution of interparticle gaps. In the jamming limit, it
consists of a delta function due to particle contacts and a background
part due to particles not in contact. As the jamming limit is approached,
it is expected that the delta-function contribution will become more
localized around contact. We derive the first exact theoretical model
for this narrowing for isostatic packings (defined below), connecting
$g_{2}$ to the probability distribution of interparticle forces $P_{f}$,
and verify the relation numerically. In this work, we present computational
data with unprecedented proximity to the jamming limit, for the first
time clearly separating the narrowing delta-function contribution
from the apparently persistent diverging background contribution.
The data show that our disordered packings have an exactly isostatic
contact network in the jamming limit, but with an unusual multitude
of nearly closed contacts. We study the properties of the contact
network and find, contrary to previous studies, no traces of polytetrahedral
packing, but rather a complex local geometry, indicating that the
geometric frustration due to the constraints of global jamming on
the local geometry is nontrivial.

\section{Theory}

A packing of $N$ hard spheres of diameter $D$ in $d$-dimensional
Euclidian space is characterized by the $(Nd)$-dimensional configuration
vector of centroid positions $\V{R}=\left(\V{r}_{1},\ldots,\V{r}_{N}\right)$.
Here we fix the center of mass of the packing (with periodic boundary
conditions), so that in fact the configuration space is of dimension
$(N-1)d$. However, we will usually neglect order unity terms compared
to $N$. The boundary conditions imposed determine the volume of the
enclosing {}``container'' $V$ and the \emph{packing} (covering)
\emph{fraction}, or \emph{density}, $\phi$.

A \emph{jammed packing} is one in which the particle positions are
fixed by the impenetrability constraints and boundary conditions \cite{Torquato_jammed,Jamming_LP}.
In particular, a packing is \emph{locally jammed} if no particle in
the system can be translated while fixing the positions of all other
particles; \emph{collectively jammed} if no subset of particles can
simultaneously be continuously displaced so that its members move
out of contact with one another and with the remainder set; and \emph{strictly
jammed} if it is collectivelly jammed and all globally uniform volume-nonincreasing
deformations of the system boundary are disallowed by the impenetrability
constraints. Assume that a configuration $\V{R}_{J}$ represents a
collectively jammed ideal packing \cite{Jamming_LP} with packing
fraction $\phi_{J}$, where there are $M$ interparticle contacts.
Next, decrease the density slightly by reducing the particle diameter
by $\D{D}$, $\delta=\D{D}/D\ll1$, so that the packing fraction is
lowered to $\phi=\phi_{J}\left(1-\delta\right)^{d}$. In this paper
we restrict ourselves to an analysis which is first order in the \emph{jamming
gap} $\delta$, $\phi\approx\phi_{J}\left(1-d\delta\right)$, and
focus on three-dimensional packings, $d=3$.

\subsection{Jamming: Configurational Trapping}

\begin{figure}
\begin{center}\includegraphics[%
  width=0.90\columnwidth,
  keepaspectratio]{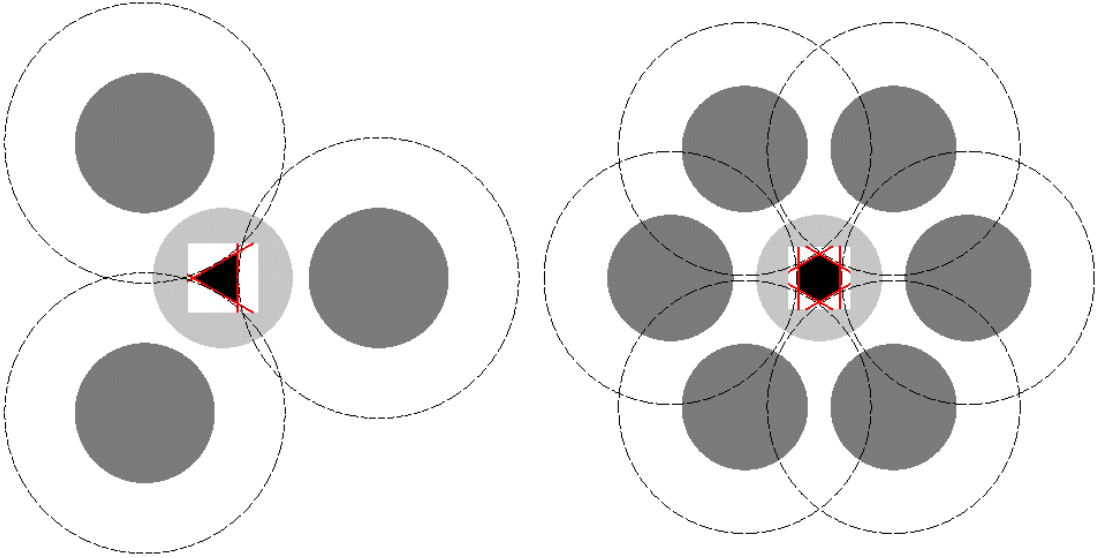}\end{center}

\caption{\label{Polytopes.2D}The polytope of allowed displacements $\mathcal{P}_{\V{\D{R}}}$
for a locally jammed disk (light shade) trapped among three (left)
or six (right, as in the triangular lattice) fixed disks (dark shade).
The exclusion disks (dashed lines) of diameter twice the disk diameter
are drawn around each of the fixed disks, along with their tangents
(solid lines) and the polytope $\mathcal{P}_{\V{\D{R}}}$ they bound
(dark). For the isostatic case on the left this polytope is a triangle
(a \emph{simplex} in two dimensions), and a hexagon for the \emph{hyperstatic}
case on the right.}
\end{figure}

It can be shown that there is a sufficiently small $\delta$ that
does not destroy the jamming property, in the sense that the configuration
point $\V{R}=\V{R}_{J}+\V{\D{R}}$ remains trapped in a small neighborhood
$\mathcal{J}_{\V{\D{R}}}$ around $\V{R}_{J}$ \cite{Connelly_Energy}.
In fact, for sufficiently small $\delta$, it can be shown that asymptotically
the set of displacements that are accessible to the packing approaches
a convex \emph{limiting polytope} (a closed polyhedron in arbitrary
dimension) $\mathcal{P}_{\V{\D{R}}}\subseteq\mathcal{J}_{\V{\D{R}}}$
\cite{FreeVolume_ClosePacked,LimitingPolytope_HS}. This polytope
is determined from the linearized impenetrability equations

\begin{equation}
\M{A}^{T}\V{\D{R}}\leq\V{\D{l}},\label{impenetrability}\end{equation}
 where $\M{A}$ is the (dimensionless) \emph{rigidity matrix}%
\footnote{This matrix combines geometrical information with the topological
connectivity information contained in the node-arc \emph{incidence
matrix} of the graph representing the contact network of the packing.
Namely, $\M{A}$ has $Nd$ rows, $d$ rows for each particle, and
$M$ columns, one for each contact. The column corresponding to the
contact between particles $i$ and $j$ is nonzero only in the rows
corresponding to the two particles, and contains the unit surface
normal vector at the point of contact \cite{Jamming_LP}.%
} of the packing, and $\V{\D{l}}$ is the set of interparticle gaps
\cite{Jamming_LP}. In our case $\V{\D{l}}=\D{D}\V{e}$, where $\M{e}$
is a vector of $M$ elements all equal to one. We can therefore focus
on the normalized polytope $\mathcal{P}_{\V{x}}:\textrm{ }\M{A}^{T}\V{x}\leq\V{e}$,
which can be scaled by a factor of $\delta D$ to obtain $\mathcal{P}_{\V{\D{R}}}$.
Examples of such polytopes for a single disk are shown in Fig. \ref{Polytopes.2D}.
A troublesome aspect, discussed in Ref. \cite{FreeVolume_ClosePacked},
is that infinite packings can never be jammed in the above sense unless
$\delta=0$, due to the appearance of unjamming mechanisms involving
collective density fluctuations. Nevertheless, computational studies
indicate that macroscopic properties derived using this polytope-based
approach do not depend on $N$, even as $N\rightarrow\infty$.

The polytope $\mathcal{P}_{\V{x}}$ is necessarily bounded for a jammed
configuration, which implies that $\M{A}$ is of full rank \cite{Jamming_LP},
and that the number of faces bounding $\mathcal{P}_{\V{x}}$, i.e.,
the number of interparticle contacts $M$, is at least one larger
than the dimensionality $d_{CS}$ of the configuration space,%
\footnote{The additional $+1$ comes because we are considering inequality constraints,
rather than equalities. One can also think of this extra degree of
freedom as representing the density, i.e., the size of the particles.
For example, looking at the left panel of Fig. \ref{Polytopes.2D}
we see that at least 3 linear inequalities are necessary to bound
a polytope in 2 dimensions.%
} $M\geq d_{CS}+1$. For collective jamming \cite{Jamming_LP} the
boundary conditions are fixed and with periodic boundary conditions
there are $d$ trivial translational degrees of freedom, so $d_{CS}=(N-1)d$.
If hard wall boundary conditions are employed then $d_{CS}=Nd$ and
one should also count contacts with the hard walls among the $M$
constraints. For strict jamming \cite{Jamming_LP} the boundary is
also allowed to deform and this introduces additional degrees of freedom.
For example, with periodic boundary conditions a symmetric non-expansive
macroscopic strain tensor is added to the configuration parameters,
giving $d_{CS}=(N-1)d+d(d-1)/2+(d-1)$ degrees of freedom%
\footnote{Here $d(d-1)/2$ gives the number of off-diagonal strain components,
and $d-1$ comes from the number of diagonal components ($d$) whose
sum is constrained to be non-positive ($-1$).%
}. \emph{Isostatic} packings are jammed packings which have the minimal
number of contacts, namely, for collective jamming\begin{equation}
M=\left\{ \begin{array}{c}
2N-1\textrm{ for $d=2$}\\
3N-2\textrm{ for $d=3$}\end{array}\right.\label{counting_collective}\end{equation}
 and for strict jamming\begin{eqnarray}
M & = & \left\{ \begin{array}{c}
2N+1\textrm{ for $d=2$}\\
3N+3\textrm{ for $d=3$}\end{array}\right.,\label{counting_strict}\end{eqnarray}
with periodic boundary conditions. Packings having more contacts than
necessary are \emph{hyperstatic}, and packings having less contacts
are \emph{hypostatic} (for sphere packings these cannot be jammed
in the above sense). For the trivial example of local jamming and
$N=1$, all particles but one are frozen in place and the free particle
must have at least $d+1$ contacts. Figure \ref{Polytopes.2D} shows
the polytope $\mathcal{P}_{\V{\D{R}}}$ for a locally jammed disk,
for both an isostatic and a hyperstatic case. In this work, we focus
on collectively jammed packings, since strictly jammed packings are
hard to produce with existing algorithms. For sufficiently large disordered
systems, the differences between collective and strict jamming are
expected to be insignificant \cite{Jamming_LP_results}.

We now consider adding thermal kinetic energy to this nearly jammed
hard-sphere packing. While the system may not be ergodic and thus
not in thermodynamic equilibrium, especially if considering disordered
packings \cite{GlassTransition_Rintoul}, one can still define a suitable
macroscopic pressure by considering only time averages as the system
executes tightly confined motion around the \emph{particular} configuration
$\V{R}_{J}$. In a sense, the configuration will explore the interior
of $\mathcal{P}_{\V{\D{R}}}$, and ergodicity is restored if one restricts
the configurational space to $\mathcal{P}_{\V{\D{R}}}$. For a finite
packing, which is sufficiently close to its jamming point, the time-averaged
properties will always be well-defined. Since the available (free)
configuration volume scales in a predictable way with the jamming
gap, $\left|\mathcal{P}_{\V{\D{R}}}\right|=\left(\delta D\right)^{Nd}\left|\mathcal{P}_{\V{x}}\right|$,
one can show that the reduced pressure is asymptotically given by
the free-volume equation of state \cite{FreeVolume_ClosePacked},\begin{equation}
p=\frac{PV}{NkT}=\frac{1}{\delta}=\frac{d}{(1-\phi/\phi_{J})}.\label{PV_free_volume}\end{equation}
 Relation (\ref{PV_free_volume}) is remarkable, since it enables
one to accurately determine the true jamming density of a given packing
even if the actual jamming point has not yet been reached, just by
measuring the pressure. We later numerically verify the validity of
Eq. (\ref{PV_free_volume}) in the vicinity of the jamming point.

\subsection{Jamming: Interparticle Forces}

As the particles travel around $\V{R}_{J}$ and the configuration
explores $\mathcal{P}_{\V{\D{R}}}$, one can average the exchange
of momentum between any two pairs of particles which share a contact
in the jammed limit (i.e., whose contact forms a face of $\mathcal{P}_{\V{x}}$),
hereafter referred to as \emph{first neighbors}, to obtain an average
interparticle \emph{force} (momentum transfer per unit time) \cite{Event_Driven_HE}.
The vector of collisional forces $\V{f}$ compares directly to the
inter-grain force networks which have been the subject of intense
experimental and theoretical study in the field of granular materials
\cite{JammedMatter_Review,Stress_Transmission,Force_Chains_Emulsions,Friction_ForceChains},
even though we are to our knowledge the first ones to directly observe
and measure them for true hard particles \cite{Event_Driven_HE}.
These forces are in local equilibrium,\begin{equation}
\M{A}\V{f}=0,\label{force_equilibrium}\end{equation}
 where we take the forces to be nonnegative%
\footnote{This sign convention is in agreement with the granular media literature,
but opposite to our own preferred notation \cite{Jamming_LP}.%
}, $\V{f}\geq0$, and normalize them to have a unit average, $\bar{f}=\V{e}^{T}\V{f}/M=1$,
in the tradition of the granular media literature. Our numerical investigations
indicate that indeed the set of time-averaged collisional forces approach
local equilibrium as the time horizon $T$ of the averaging increases,
in a inverse-power law manner, $\left\Vert \M{A}\V{f}\right\Vert \sim T^{-1}$.
We can therefore obtain interparticle forces relatively accurately
given sufficiently long molecular dynamics runs. While Eq. (\ref{force_equilibrium})
will have a unique solution if and only if the contact network of
the packing is isostatic, even for hyperstatic packings, such as the
FCC packing, the equilibrium set of forces should be unique. In fact,
one can prove that the force between two particles will be proportional
to the surface area of the face of $\mathcal{P}_{\V{x}}$ formed by
the contact in question. 

It is interesting to observe that if one has an \emph{arbitrary} point
$\V{\D{R}}\in\mathcal{P}_{\V{\D{R}}}$, the interparticle gaps due
to nonzero jamming gap will be $\V{\D{l}}\approx\M{A}^{T}\V{\D{R}}-\delta D\V{e}$,
so that\begin{equation}
\V{f}^{T}\V{\D{l}}\approx(\M{A}\V{f})^{T}\V{\D{R}}-MD\delta=-MD\delta.\label{f_to_delta}\end{equation}
Eq. (\ref{f_to_delta}) enables one to determine how far from the
jamming density a packing is without actually reaching the jamming
point. This can be a useful alternative to using Eq. (\ref{PV_free_volume})
when the hard-sphere pressure is not available, but interparticle
forces are, such as, for example, with packings generated by algorithms
using stiff {}``soft'' spheres \cite{Quenching_Jamming}.

As already pointed out, hypostatic packings cannot be jammed. However,
it is possible for a hypostatic packing to be \emph{locally maximally
dense}, in the sense that no continuous motion of the particles can
increase the density to first order. In other words, the particles
must first move and unjam (which must be possible for a hypostatic
sphere packing) before the density can increase. In particular, a
packing of contacting particles for which a set of interparticle forces
$\V{f}$ in equilibrium exists, is locally maximally dense. In a sense,
the interparticle forces resist further increase of the density. As
we discuss later, our packing generation algorithm sometimes terminates
with such packings since it tries to continually increase the density.

\subsection{Pair Correlation Function Around Contact}

We now turn to the central subject of this work: The shape of the
(orientation-averaged) pair correlation function $g_{2}(r)$ for small
jamming gaps. In particular, we will focus on interparticle distances
$r$ that are very close to $D$. We express $g_{2}(l)$ in terms
of the nonnegative \emph{interparticle gaps} $l=r-D$. The polytope
picture above says that only the $M$ first-neighbor particle pairs
will contribute to the shape of $g_{2}(l)$ right near contact, i.e.,
for gaps up to $l_{\textrm{max}}$, where $l_{\textrm{max}}$ is the
largest distance from the centroid of $\mathcal{P}_{\V{\D{R}}}$ to
one of its faces. This contribution will become a delta function in
the jamming limit. Particle pairs not in contact will not contribute
to $g_{2}(l)$ until gaps larger than the minimal further-neighbor
gap $l_{FN}$, and for now we will implicitly assume that $l_{FN}\gg l_{\textrm{max}}$,
so that there is a well-defined \emph{delta-function region} $g_{2}^{(\delta)}(l)\equiv g_{2}(l\ll l_{FN})$.
This delta function region has previously been investigated theoretically
for crystal packings, primarily \cite{LimitingPolytope_HS}. In this
work, we derive exact theoretical expressions for this region for
isostatic packings, as well as numerically study vacancy-diluted FCC
crystals and partially crystallized packings.

\subsubsection{Isostatic Packings}

We first focus on the probability distribution for observing an interparticle
gap $l$, $P_{l}(l)$, which is related to $g_{2}^{(\delta)}(l)$
via a simple normalization factor. The contribution $\tilde{P}(l)$
from a specific contact is determined from the area $\tilde{S}(l)$
of the cross section of $\mathcal{P}_{\V{x}}$ with a plane parallel
to the face corresponding to the contact and at a distance $l$ from
the face, $\tilde{P}(l)\sim\tilde{S}(l)$ \cite{LimitingPolytope_HS}.
The critical observation we make is that for an isostatic contact
network, $\mathcal{P}_{\V{x}}$ is a \emph{simplex} and thus immediately
we get $\tilde{S}(l)\sim\left[(h-l)/h\right]^{M}$, where $h$ is
the height of the simplex corresponding to this particular face, $h=M\left|\mathcal{P}_{\V{x}}\right|/S$,
$S=\tilde{S}(0)$. After normalization of $\tilde{P}(l)$ and averaging
over all interparticle contacts, we obtain that\[
P_{l}(l)=\int_{h=l}^{\infty}\frac{M}{h}\left[1-\frac{l}{h}\right]^{M}P_{h}(h)dh,\]
which shows that if we know the distribution $P_{h}$ of heights for
the simplex $\mathcal{P}_{\V{x}}$, or equivalently, the distribution
of surface areas $S$ of the faces of the polytope $P_{S}(S)$, we
would know $P_{l}$ and thus $g_{2}^{(\delta)}$.

Since the interparticle force $f\sim S$, we see immediately that
the distribution of face areas is equivalent to the distribution of
interparticle forces $P_{f}(f)$, and in fact it is easy to derive
that

\[
h/\D{D}=1+\frac{\V{e}^{T}\V{f}}{f}\approx\frac{M}{f},\]
which gives in the limit $M\rightarrow\infty$\begin{eqnarray*}
P_{l}(l)=\int_{f=0}^{M/l}\frac{f}{\D{D}}\left[1-\frac{lf}{M\D{D}}\right]^{M}P_{f}(f)df & \approx\\
\approx\frac{1}{\D{D}}\int_{0}^{\infty}fP_{f}(f)\exp(-fl/\D{D})df & = & \frac{1}{\D{D}}\mathcal{L}_{l/\D{D}}\left[fP_{f}(f)\right],\end{eqnarray*}
 where $\mathcal{L}_{s}$ denotes the Laplace transform with respect
to the variable $s$. We have the normalization condition $\int_{0}^{\infty}P_{l}(l)dl=1$
and additionally\[
DP_{l}(0)=\frac{D}{\D{D}}\int_{0}^{\infty}fP_{f}(f)df=\frac{D}{\D{D}}=p.\]
If we now relate $P_{l}(l)$ to $g_{2}^{(\delta)}(l)$,\[
g_{2}^{(\delta)}(l)=\frac{2MV}{4\pi D^{2}N^{2}}P_{l}(l)=\frac{\bar{Z}D}{24\phi}P_{l}(l),\]
where $\bar{Z}=2M/N=2d=6$ is the \emph{mean coordination number},
we obtain the central theoretical result\begin{equation}
g_{2}^{(\delta)}(l)=\frac{p}{4\phi}\mathcal{L}_{l/\D{D}}\left[fP_{f}(f)\right].\label{g2_from_P_f}\end{equation}

\subsection{Classification of Jammed Packings}

\begin{figure}
\begin{center}\includegraphics[%
  width=0.60\textwidth]{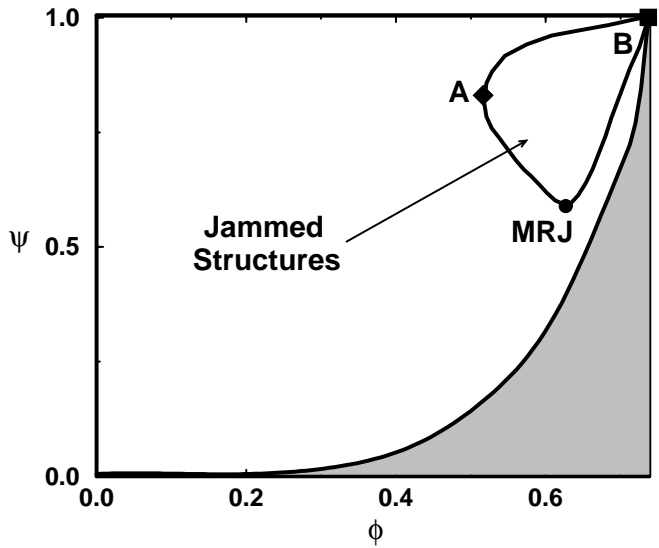}\end{center}

\caption{\label{Phi_Psi_Jammed}A highly schematic plot of the subspace in
the density-disorder ($\phi-\psi$) plane where strictly jammed three-dimensional
packings exist. Point $A$ corresponds to the lowest-density jammed
packing, and it is intuitive to expect that a certain ordering will
be needed to produce low-density jammed packings. Point $B$ corresponds
to the most dense jammed packing. Point MRJ represents the maximally
random jammed state. This is the most disordered jammed packing in
the given jammed category (locally, collectively or strictly jammed).
We conjecture that the Lubachevsky-Stillinger packing algorithm \cite{LS_algorithm,LS_algorithm_3D}
typically produces packings along the right (maximally dense) branch,
and we do not know of an algorithm that produces packings along the
left (minimally dense) branch.}
\end{figure}

Jammed hard-sphere packings can be classified based on their density
$\phi$. However, such a classification is clearly not sufficient
in order to distinguish between \emph{ordered} and \emph{disordered}
(often called random, despite the shortcomings of such terminology)
packings \cite{Torquato_MRJ,Anu_order_metrics}. In fact, packings
can have various degrees of order in them, and for hard-sphere packings
the dominant form of ordering is crystallization into variants of
the FCC lattice. We can use a hypothetical scalar order-metric $\psi$
to measure the amount of order in a packing, such that $\psi=1$ corresponds
to fully ordered (for example, the perfect FCC crystal), and $\psi=0$
corresponds to perfectly disordered (Poisson distribution of sphere
centers) packings. Very large jammed packings are thus classified
based on their position in the density-disorder ($\phi-\psi$) plane,
as sketched in Fig. \ref{Phi_Psi_Jammed}, as taken from Ref. \cite{Torquato_MRJ}.
A state of special interest is the MRJ state, representing the collection
of \emph{maximally random jammed} packings, believed to be closely
related to the traditional but ill-defined concept of random close
packing (RCP) in three dimensions if strict jamming is considered,
and to have a density of about $\phi\approx0.64$ in three dimensions%
\footnote{Contrary to popular belief, the traditional concept of RCP does not
have a two-dimensional analog for monodisperse disks \cite{Jamming_LP_results,Comment_OHern}.%
}. Additionally, the perfect FCC crystal and variants thereof correspond
to the most dense jammed packing, with $\phi\approx0.74$. This work
will focus on these two points in the $\phi-\psi$ plane. However,
it is possible to produce packings with intermediate amounts of order
and densities, for example, by allowing partial crystallization.

\section{Computational Results}

We use event-driven molecular dynamics \cite{Event_Driven_HE} as
the primary computational tool for our investigations. This enables
us to perform exact molecular dynamics on hard-particle packings very
close to the jamming point, which is not possible with traditional
time-driven molecular dynamics algorithms. The algorithm monitors
a variety of properties during the computational run, including the
{}``instantaneous'' pressure, as calculated from the total exchanged
momentum in all interparticle collisions during a certain short time
period $\D{t}$. By allowing the shape of the particles to change
with time, for example, by having the sphere diameter grow (shrink)
uniformly at a certain (possibly negative) expansion rate $dD/dt=2\gamma$,
one can change the packing density. If the change is sufficiently
slow, the system will be in approximate (metastable) equilibrium during
the densification, and one can rather effectively gather quasi-equilibrium
data as a function of density.

Event-driven molecular dynamics in which the particles (quickly) grow
in size in addition to their thermal motion at a certain expansion
rate, starting from a random (Poisson) distribution of points, produces
a jammed state with a diverging collision rate. This is the well-known
Lubachevsky-Stillinger (LS) packing algorithm \cite{LS_algorithm,LS_algorithm_3D},
which we have used and modified \cite{Event_Driven_HE} to generate
all the disordered hard-sphere packings for this study. During the
initial stages, the expansion has to be fast to suppress crystallization
and maximize disorder \cite{Anu_order_metrics}, and delaying further
discussion to later sections, we will assume that the disordered packings
used in this study are representative of the MRJ state. It is important
to note that the algorithm typically produces packings that have \emph{rattling}
particles, i.e., particles that do not have true contacts with particles
in the jammed \emph{backbone} of the packing, and can be removed without
affecting the jamming category of the final packing. We will discuss
procedures for identification of such rattlers in what follows.

\begin{figure}
\begin{center}\includegraphics[%
  width=0.75\columnwidth,
  keepaspectratio]{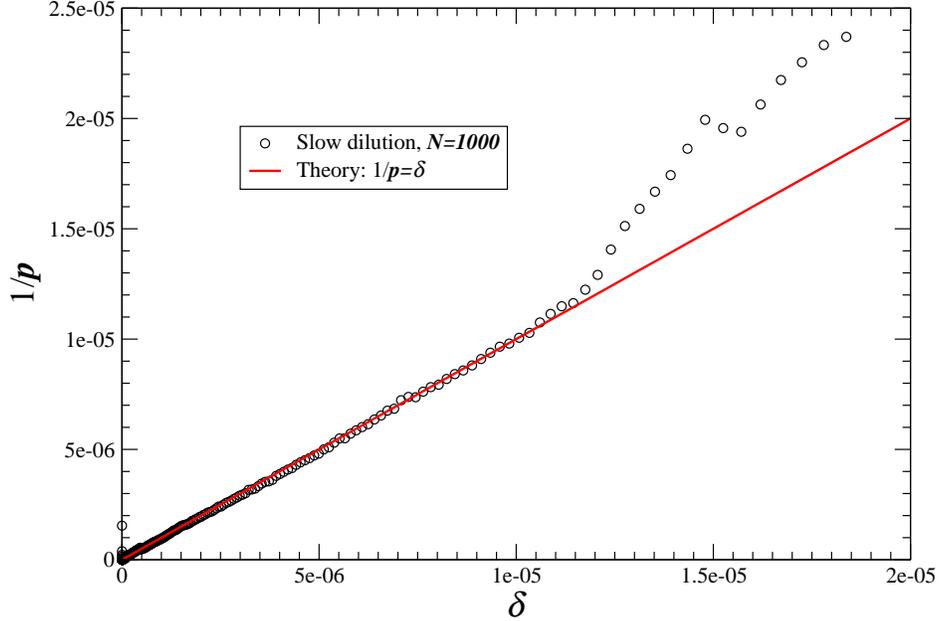}\end{center}

\caption{\label{PV.HS.N=3D1000.dilution}The inverse of the {}``instantaneous''
(averaged over several hundred collisions per particle) pressure of
a nearly jammed (isostatic) packing of 1000 particles, as it is slowly
diluted (using a negative expansion rate for the particles in the
molecular dynamics algorithm $\gamma=10^{-5}$) from $\phi_{J}\approx0.627$
until an unjamming particle rearrangement occurs. Up to this occurrence,
the free-volume theoretical relation $p=\delta^{-1}$ is satisfied
to very high accuracy. Rattlers have been removed from the packing.}
\end{figure}

\begin{figure}
\begin{center}\includegraphics[%
  width=0.75\columnwidth,
  keepaspectratio]{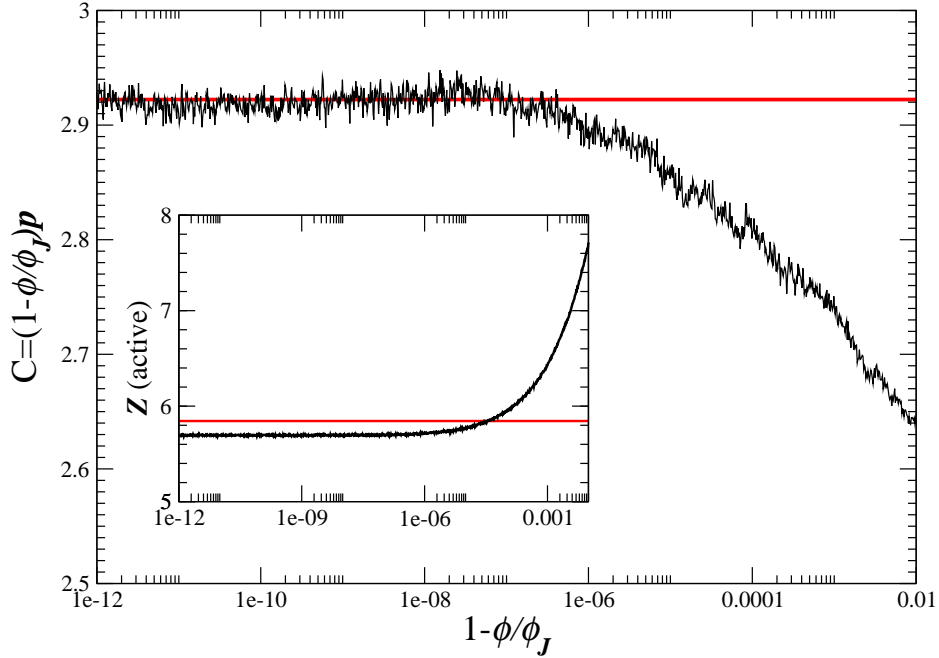}\end{center}

\caption{\label{PV.HS.coefficient}The coefficient $C$ during a typical slow
densification (expansion rate is $10^{-4}$) of a $10,000$-particle
system, starting from an equilibrated liquid at $\phi=0.5$ up to
jamming. The final packing has $259$ rattlers, so the expected coefficient
is $3\cdot0.9741\approx2.92$, a value which is shown with a red line.
It is clear that close to the jamming point Eq. (\ref{PV_free_volume})
is very accurate, but a marked lowering from a coefficient of $3$
is seen for pressures lower than about $10^{6}$, likely explaining
the coefficient $2.67$ reported in works of Speedy \cite{QuenchRate_Speedy,Metastable_Fluid_HS}.
The inset shows the estimated {}``collisional'' coordination, defined
as the average number of different particles that a particle has collided
with during a time interval of about 100 collisions per particle,
during the same densification. The expected number $6\cdot0.9741\approx5.85$
is shown (this number is not asymptotically reached exactly since
some of the $M$ contacts do not participate in collisions frequently
enough to be registered during the time interval used), and we see
that as many as $8$ contacts per particle are active at sufficiently
large jamming gaps.}
\end{figure}

To our knowledge, no verification of the exactness of Eq. (\ref{PV_free_volume})
for disordered packings exists in the literature. The perfect FCC
crystal is stable until rather low densities, and the pressure seems
to be rather accurately predicted by the free-volume approximation
in a wide range of densities around close packing. This has been observed
in the literature and a suitable corrective term was determined \cite{Solid_Branch_HS}.
However, for disordered packings, previous studies have identified
a coefficient smaller than $3$ in the numerator, namely $2.67$ \cite{QuenchRate_Speedy,Metastable_Fluid_HS}.
In Fig. \ref{PV.HS.N=3D1000.dilution}, we numerically verify the
validity of Eq. (\ref{PV_free_volume}) with very high accuracy for
disordered packings. In Fig. \ref{PV.HS.coefficient}, we show the
change of the coefficient (the configurational heat capacity in units
of $Nk$) $C=(1-\phi/\phi_{J})p$ with density. Agreement with the
theoretical $C=d=3$ is observed sufficiently close to the jamming
point, but with rapid lowering of the coefficient from $3$ away from
the jamming point. This is because for sufficiently large jamming
gaps, contacts other than the $M$ true contacts start contributing
to the collisions, and the polytope-based picture we presented so
far does not apply exactly. We demonstrate this in Fig. \ref{PV.HS.coefficient}
by showing the number of contacts which participate in collisions
(\emph{active contacts}) as the jamming point is approached. Our investigations
indicate that previous studies did not examine at the range of densities
appropriate for the theory presented above and did not properly account
for the rattlers.

\subsection{Disordered Packings}

\begin{figure}
\begin{center}\includegraphics[%
  width=0.75\columnwidth,
  keepaspectratio]{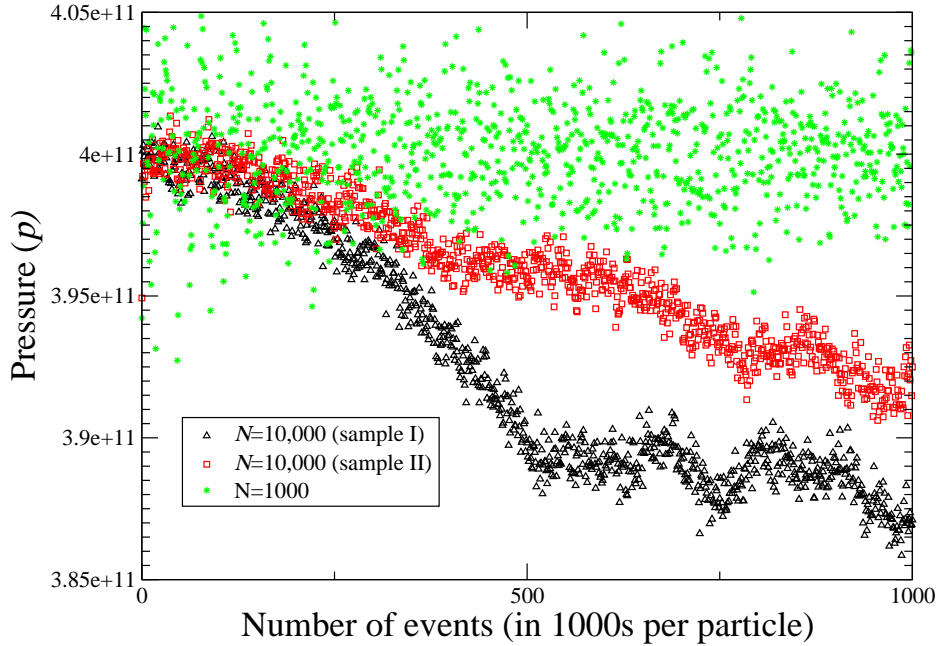}\end{center}

\caption{\label{PressureLeak.HS}The short-term ({}``instantaneous'') pressure
versus number of events (mostly binary collisions) processed by the
molecular dynamics algorithm \cite{Event_Driven_HE}, corresponding
to a total run of about half a million collisions per particle. For
the 1000-particle packing the pressure is stable, but for the larger
packings a systematic pressure leak is observed.}
\end{figure}

We have verified in previous publications that LS packings are typically
collectively jammed \cite{Jamming_LP_results} using a testing procedure
based on linear programming \cite{Jamming_LP}. Unfortunately, the
linear programming library used in the implementation cannot really
achieve the kind of numerical accuracy that we require in this work,
specifically that for packings which are jammed almost to within full
numerical precision ($\delta=10^{-15}-10^{-12}$). Additionally, it
cannot handle three-dimensional packings of more than about a thousand
particles. Another test for jamming, which we have found to be reliable
for the purposes of this work, is to take the final packing produced
by the LS procedure and then run standard event-driven molecular dynamics
on it for long periods of time (on the order of thousands to hundreds
of thousands of collisions per particle) and monitor the {}``instantaneous''
pressure. If the packing is jammed, this pressure will be stable at
its initial value. However, if the packing is not truly jammed, we
have observed that the pressure slowly decays with time, the slower
the {}``pressure leak'' the more {}``jammed'' the initial packing
is, as illustrated in Fig. \ref{PressureLeak.HS}. Similar observations
are made in Ref. \cite{QuenchRate_Speedy}. In addition, we track
the average particle displacement (from the initial configuration)
and check to see if there is a systematic drift with time away from
the initial configuration. The two tests always agreed: A pressure
leak always corresponds to a systematic drift away from the initial
configuration.

We have observed that LS packings densified to within numerical capability
only pass this rigorous jamming test of having no pressure leak if
during the final stages of the LS densification the expansion rate
is very small compared to the average thermal velocity (maintained
constant via a velocity rescaling thermostat \cite{Event_Driven_HE})
of the particles (about five orders of magnitude or less). Similar
observations are made in Ref. \cite{QuenchRate_Speedy}. If the expansion
rate is too fast, we have found that the packings jam in slightly
hypostatic configurations, where there are not enough particle contacts
to ensure jamming. In particular, some particles have 2 or 3 contacts
(and of course rattlers are present). In order for a set of balanced
forces to exist (which as we discussed is a necessary condition for
a packing to be locally maximally dense) when a particle has less
than $4$ contacts, these contacts must be in a degenerate geometric
configuration, namely 3 coplanar or 2 collinear contacts. We have
indeed verified that this is what happens in the hypostatic packings
produced by the LS algorithm. The number of such geometric peculiarities
increases with increasing expansion rate, and also for more ordered
packings, as we discuss later.

\begin{figure}
\begin{center}\includegraphics[%
  width=0.90\columnwidth,
  keepaspectratio]{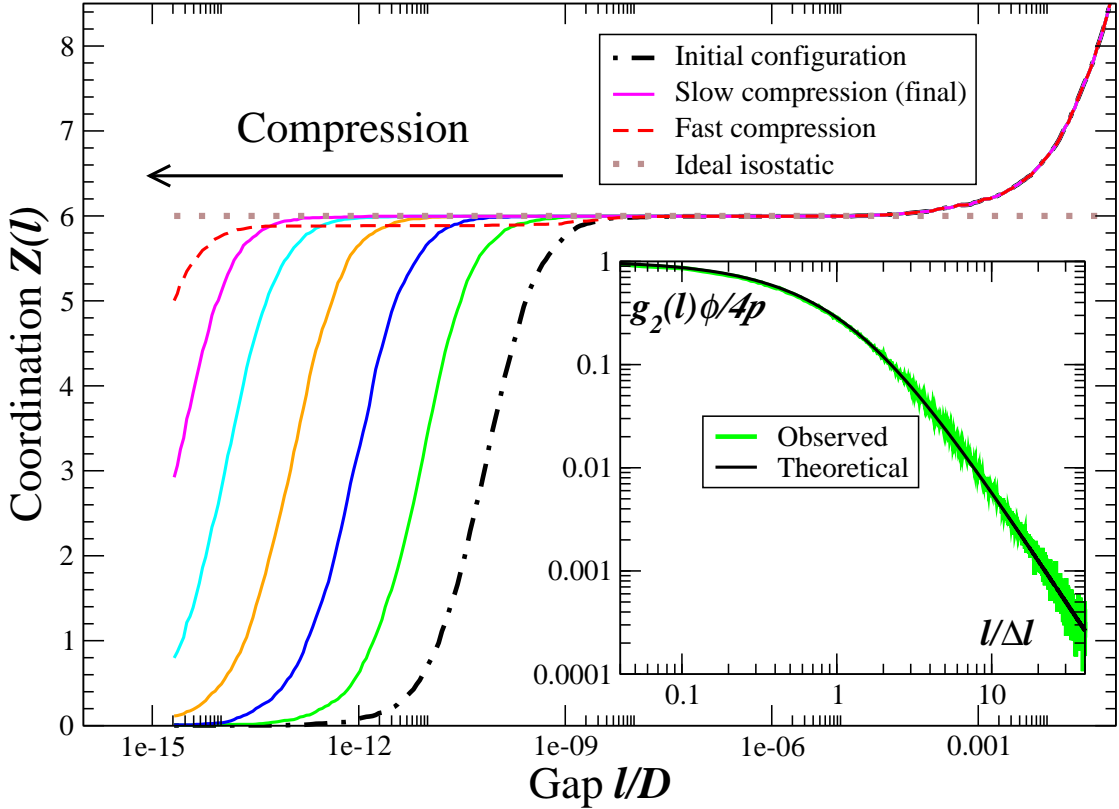}\end{center}

\caption{\label{G2.HS.compression}The cumulative coordination $Z(l)$ (i.e.,
the integral of $g_{2}(l)$) as a function of the gap tolerance $l$,
for a sequence of snapshots of a 1000-particle packing during the
final compression stages of the LS algorithm. For a sufficiently slow
expansion (expansion rate is $10^{-5}$ times the average thermal
velocity), the packing is clearly seen to jam in an isostatic configuration.
A subisostatic configuration is found for fast expansion (expansion
rate is comparable to the thermal velocity). The inset shows the properly
normalized derivative of $Z(l)$, right around contact, along with
a comparison to our semi-theoretical prediction for $g_{2}^{(\delta)}(l)$,
for a packing with $\delta=2.5\cdot10^{-12}$.}
\end{figure}

We illustrate the progress of the densification during the final stages
of the algorithm in Fig. \ref{G2.HS.compression}. The figure shows,
for several snapshots of the packing during the densification, the
\emph{cumulative coordination number} \[
Z(l)=\frac{N}{V}\int_{r=D}^{D+l}4\pi r^{2}g_{2}(r)dr=24\phi\int_{r=D}^{D+l}\left(\frac{r}{D}\right)^{2}g_{2}(r)\frac{dr}{D},\]
i.e., the average number of particles within a gap $l$ from a given
particle. We we will often use this quantity instead of $g_{2}(l)$.
For the first time in the vast literature on random packings, a clear
separation is seen between the delta-function contribution $Z^{(\delta)}(l)$,
which becomes more localized around contact, and the background increase
in the mean coordination from the isostatic contact value of $\bar{Z}=6$,
which remains relatively unaffected by the densification. For small
packings ($N=1000$), the value of $Z(l)$ is fixed at 6 for a remarkably
wide range of gaps, as much as 9 orders of magnitude for the final
packings. Fast densification is seen to lead to subisostatic packings
in Fig. \ref{G2.HS.compression}, leaving a certain fraction of the
contacts {}``open''. Stopping the expansion invariably leads to
a decay of the macroscopic pressure for such subisostatic packings.

By using heuristic strategies, we were able to find (slow) densification
schemes which produced packings which are indeed ideally jammed within
almost full numerical precision, at least for packings of $N=1000$
particles or less. In fact, the plateau in $Z(l)$ was at exactly
(up to a single contact!) an isostatic number of contacts, $M=3N-2$,
for all the packings produced via a carefully guided LS algorithm.
It is essential that here $N$ is the number of particles in the jammed
backbone of the packing \cite{Jamming_LP}, i.e., rattlers \cite{LS_algorithm_3D}
with fewer than $2$ contacts have been removed from the packing.
It seems that the algorithm produces packings with about $2.2\%$
rattlers, and so the density of the disordered packings we look at
is typically $\phi\approx0.625-0.630$, rather than the widely known
$\phi\approx0.64$. Despite a concentrated effort and lots of expended
CPU time, we have been unable to achieve true isostaticity for $10,000$-particle
packings. This is illustrated in Fig. \ref{PressureLeak.HS}, where
it is clearly seen that the pressure in the large packings does not
remain constant over long periods of time (about a million collisions
per particle). It is therefore not strictly justified to consider
these packings within the framework of ideal jammed packings that
we have adopted here. However, it is readily observed that over finite
and not too long time intervals (for example several thousands of
collisions per particle), the large packings conform to the predictions
of the theory developed here. In particular, the collisional forces
form a balanced force network with essentially the same $P_{f}(f)$
as the truly jammed smaller packings, and the pressure is given by
Eq. (\ref{PV_free_volume}) with very high accuracy, where $\delta$
can be determined, for example, via Eq. (\ref{f_to_delta}). We therefore
believe it is justified to use the larger packings for certain analysis
where better statistics are needed.

The main goal of this work is to explore and explain Fig. \ref{G2.HS.compression},
and in particular, to investigate both the {}``delta-function'',
or \emph{contact}, contribution $g_{2}^{(\delta)}$, which should
integrate to produce the isostatic average coordination $\bar{Z}=2M/N=6$,
and the {}``background'' or \emph{near-contact} $g_{2}^{(b)}$,
for gaps from about $100\delta D$ to $10^{-1}D$. This latter one
has already been observed in an experimental study of hard spheres
\cite{Tomography_SpherePackings}, and in computational studies of
stiff {}``soft'' spheres \cite{Friction_Packing,Quenching_Jamming}.
These various studies find a nearly square-root divergence, $g_{2}^{(b)}(l)\sim1/\sqrt{l}$,
and Ref. \cite{Friction_Packing} observes that this is an integrable
divergence and thus clearly separate from the delta function. Our
results, shown in Fig. \ref{G2.HS.compression}, are the first unambiguous
and precise separation of the two pieces of the pair correlation function
around contact near jamming. Our numerical data has precision ($\delta<10^{-13}$)
not previously attained, since such proximity to ideal jammed hard-sphere
packings can only be achieved in a true hard-sphere algorithm, and
at present only event-driven molecular dynamics seems to provide the
required numerical robustness. It is rather interesting that although
graphs showing the hard-sphere $g_{2}(l)$ in the literature have
clearly demonstrated a divergence in $g_{2}(l)$ near contact for
at least three decades \cite{RCP_SerialDeposition}, this seems to
never have been clearly documented or investigated. We are led to
believe that researchers were under the false impression this divergence
is a signature of the delta-function contribution, and thus expected
it to further narrow and disappear at true jamming.

\subsubsection{Delta-Function (Contact) Contribution}

\begin{figure}
\begin{center}\includegraphics[%
  width=0.95\columnwidth,
  keepaspectratio]{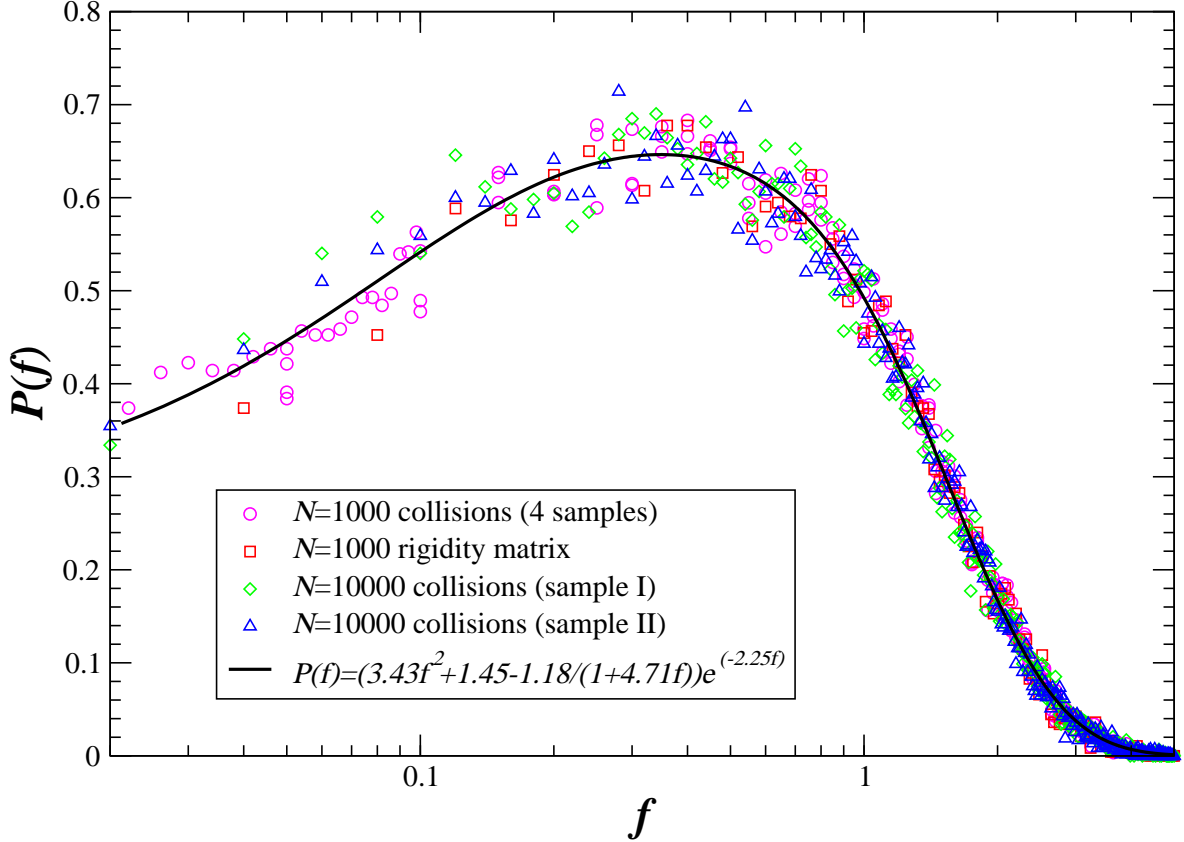}\end{center}

\caption{\label{P_f.HS.fits}Computational data on the interparticle force
distribution along with the best fit we could achieve. Packings of
both $1,000$ and $10,000$ particles, using either molecular dynamics
to average the collisional forces, or inversion of the rigidity matrix,
were used, consistently producing the same probability distribution.
Comparison to other data in the granular-media literature is beyond
the scope of this work.}
\end{figure}

We first verify that our theory correctly predicts the shape of $g_{2}^{(\delta)}(l)$.
In order to verify relation (\ref{g2_from_P_f}) numerically, a form
for $P_{f}(f)$ is needed. Force networks in particle packings have
been the subject of intense theoretical and experimental interest
\cite{JammedMatter_Review,Friction_ForceChains,ForceChains_OHern,Force_Chains_Emulsions,ForceChains_Science},
and it has been established that $P_{f}$ decays exponentially at
large forces for a variety of models. The behavior of $P_{f}$ for
small forces has not been agreed upon, the central question being
whether the infinite-system-limit $P_{f}(0)$ is nonzero. No theoretical
model has been offered yet that truly answers this question. Part
of the difficulty is that the answer likely depends not only on the
system in question, but also on the definition of $f$. In a true
ideal collectively jammed isostatic packing, which is necessarily
finite, all interparticle forces, \emph{must} be strictly positive,
and in fact are determined uniquely through Eq. (\ref{f_direct_inversion}),\begin{equation}
\V{f}=\left[\begin{array}{c}
\M{A}\\
\V{e}^{T}\end{array}\right]^{-1}\left[\begin{array}{c}
\V{0}\\
1\end{array}\right],\label{f_direct_inversion}\end{equation}
without any mention of interparticle potentials or influence of external
fields or loads like gravity, or thermal dynamics. The limiting probability
distribution of these interparticle forces as the packing becomes
larger, if it exists, can be positive at the origin, indicating that
finite but large packings have limiting polytopes with a few extremely
small faces, or equivalently, are very elongated along certain directions.
We have numerically studied the form of $P_{f}(f)$ for almost jammed
random packings of $N=1000$ and $N=10,000$ spheres by using molecular
dynamics to observe the collisional forces between first neighbors,
and also by directly using Eq. (\ref{f_direct_inversion}) for the
smaller packings%
\footnote{Efficiently inverting the rigidity matrix for very large three dimensional
packings is a rather challenging numerical task which we have not
yet tackled.%
} (this offers better accuracy for small forces). The results are shown
in Fig. \ref{P_f.HS.fits}. We clearly see a peak in $P(f)$ for small
forces, as observed in the literature for jammed packings of soft
particles \cite{ForceChains_OHern}, and it appears that there is
a finite positive probability of observing zero interparticle force.
We will return to this point later.

The observed $P_{f}(f)$ can be well fitted for medium and large forces
by $P_{f}(f)=(Af^{2}+B)e^{-Cf}$, with a small correction needed to
fit the small-force behavior, as used in Fig. \ref{P_f.HS.fits}.
This small correction has a negligible impact on the Laplace transform
of $fP_{f}(f)$, and in fact a very good approximation to $g_{2}^{(\delta)}(l)$
in Eq. (\ref{g2_from_P_f}) is provided by just using\begin{equation}
\mathcal{L}_{x}\left[fP_{f}(f)\right]=\frac{6A}{\left(x+C\right)^{4}}+\frac{B}{\left(x+C\right)^{2}}.\label{Laplace_transform}\end{equation}
In the inset in Fig. \ref{G2.HS.compression}, we show a comparison
between the $g_{2}^{(\delta)}(l)$ we observe computationally and
the one given by Eqs. (\ref{g2_from_P_f}) and (\ref{Laplace_transform})
and the empirical fit to $P_{f}(f)$ in Fig. \ref{P_f.HS.fits}. An
essentially perfect agreement is observed. Our focus here is on small
forces, however we do wish to note that our data cannot confidently
rule out a Gaussian component to $P_{f}$ for large forces and that
a slight quadratic component does seem to be visible when $P_{f}(f)$
is plotted on a log-log plot.

\subsubsection{Near-Contact Contribution}

\begin{figure}
\begin{center}\includegraphics[%
  width=0.90\columnwidth,
  keepaspectratio]{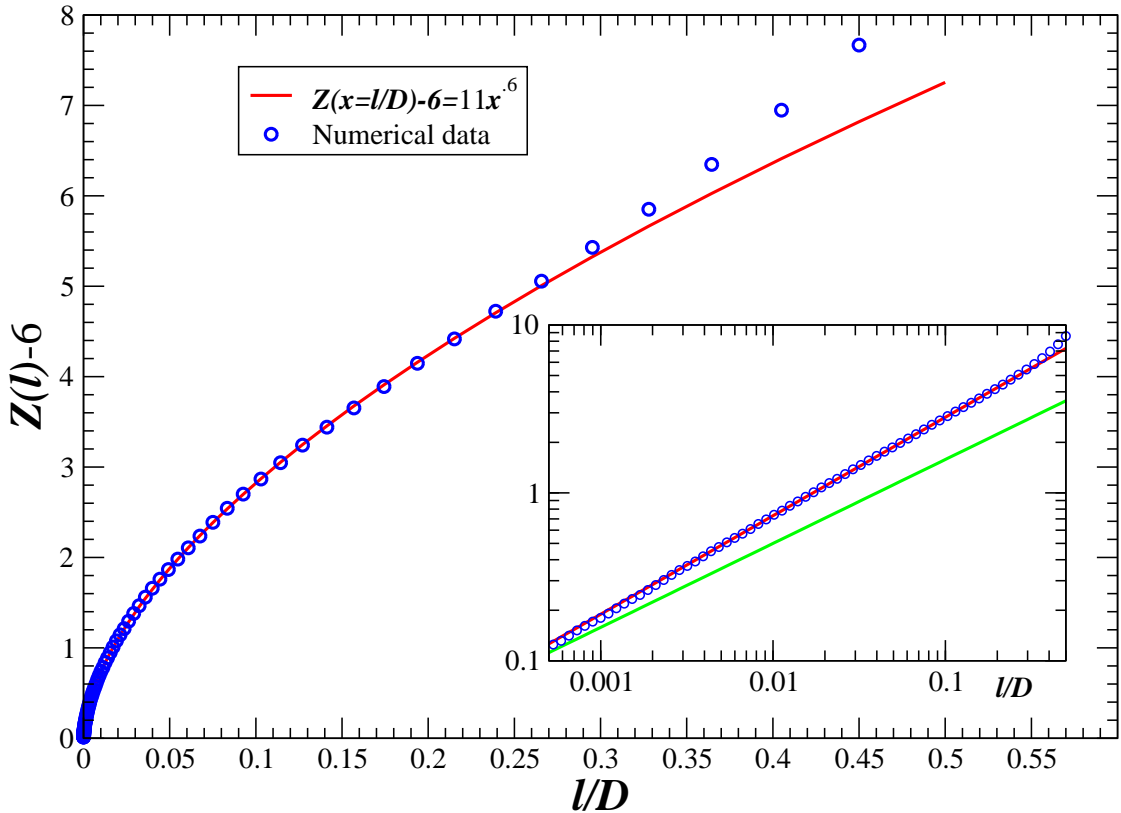}\end{center}

\caption{\label{G2.HS.near-contacts}The near-contact $Z^{(b)}(l)$ for a
nearly-jammed $10,000$-particle packing, along with a power law fit
for small gaps, shown in both a linear-linear scale and a log-log
scale (inset). In this inset we also show a line with slope $0.5$
(i.e., a square root dependence), which is clearly inconsistent with
the numerical data.}
\end{figure}

In Fig. \ref{G2.HS.near-contacts} we investigate the near-contact
contribution to $g_{2}(l)$. We have found that $Z^{(b)}(l)$ has
a power law behavior over a surprisingly wide range of gaps, up to
the first minimum of $g_{2}$ at $l\approx0.25D$, $Z^{(b)}(l)\approx11(l/D)^{0.6}$,
as shown in the figure. Note that this range is too wide for\[
g_{2}^{(b)}(x)=\frac{1}{24\phi(1+x)^{2}}\frac{dZ^{(b)}(x)}{dx}\]
 to be a perfect power law, where $x=l/D$, as used to fit numerical
data in other studies (which have not investigated nearly as wide
a range of gaps as we do here) \cite{Friction_Packing,Tomography_SpherePackings}.
Additionally, the observed exponent is clearly distinguishable from
an inverse square root divergence in $g_{2}^{(b)}(l)$, as proposed
in the literature \cite{Friction_Packing}. 

We do not have a theoretical explanation for this functional behavior
of $Z^{(b)}(l)$, however, the remarkable quality of the fit in Fig.
\ref{G2.HS.near-contacts} hints at the possibility of a (simple)
scaling argument. Some simple observations can be made by assuming
that \begin{equation}
Z(x)=\bar{Z}+ax^{1-\alpha}\textrm{ for $0<x\leq\beta$},\label{Z_power_law}\end{equation}
where $\alpha$ is an exponent $0\leq\alpha\leq1$, and $\beta<1$
determines the extent of this power-law dependence. The corresponding
pair correlation function of course exhibits an inverse power-divergence
with exponent $\alpha$, except when $\alpha=1$, when it is identically
zero%
\footnote{Note that $g_{2}^{(b)}(x)$ cannot have a simple-pole divergence since
this would lead to a logarithmic divergence in $Z^{(b)}(x)$, which
must be finite for all finite $x$.%
}. The exponent $\alpha$ clearly will depend on the amount of order
present in the packing, i.e., the position of the packing in the density-order
diagram of Fig. \ref{Phi_Psi_Jammed}. We expect that it will increase
with increasing order, since $\alpha\rightarrow0$ would indicate
a constant $g_{2}(l)$ near contact, a signature of the ideal gas,
while $\alpha\rightarrow1$ would indicate a clear distinction between
the first and second shell of neighbors (i.e., a wide range of gaps
with very few contacts) typical of crystal packings. Under the assumption
that a power-law divergence in $g_{2}$ is appropriate, an intermediate
value of $\alpha$ between $0$ and $1$, as we find numerically,
is therefore expected. Some bounds on the range of possible $\alpha$
can be obtained from bounds on $Z(x)$ derived from geometric constraints
(for example, $Z(x)<13$ for a certain range of $x$ since the sphere
kissing number is $12$ in three-dimensions), but the exact value
is not simple to predict%
\footnote{The three parameters $\alpha$, $\beta$ and $a$ are thus not independent
of one another. For example, requiring that $g_{2}^{(b)}(x)>1$ and
$Z(x)<12$ for $0<x\leq\beta$ gives the weak constraints $a(1-\alpha)>24\phi\beta^{2}(\beta-1)^{\alpha}$
and $a(\beta-1)^{1-\alpha}<12-\bar{Z}$.%
}.

\subsubsection{Away from Contact: Split Second Peak}

\begin{figure}
\begin{center}\includegraphics[%
  width=0.90\columnwidth,
  keepaspectratio]{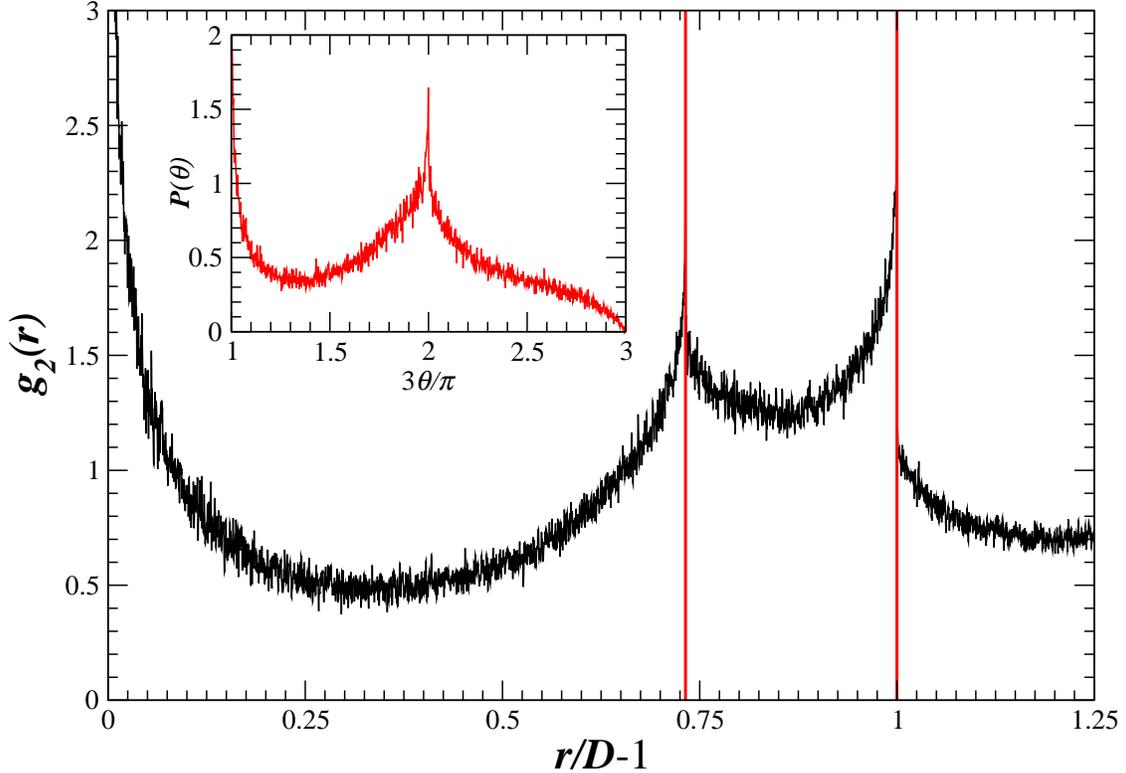}\end{center}

\caption{\label{g2.HS.split-peak}Computational data on the split second peak
of $g_{2}(r)$ averaged over $5$ packings of $10,000$ particles.
The values $r=\sqrt{3}D$ and $r=2D$ are highlighted, and match the
two observed discontinuities. Also visible is the divergence near
contact. The inset shows the probability distribution $P_{\theta}(\theta)$
of bond-pair angles in the contact network of the packings, also revealing
two divergences at $\theta=\pi/3$ and at $\theta=2\pi/3$. No peaks
are observed at $r=\sqrt{2}D$ or $r=\sqrt{5}D$, which are typical
of crystal packings, indicating that there is no detectable crystal
ordering in the packing. }
\end{figure}

Although the primary focus of this work is on the behavior of $g_{2}(r)$
around contact, it is instructive to also look at the split second
peak of the pair correlation function, shown for a sample of packings
of $10,000$ particles in Fig. \ref{g2.HS.split-peak}. Only two clear
discontinuities are seen, one at exactly $r=\sqrt{3}D$, and one at
$r=2D$. The latter is very clearly asymmetrical, with a sharp decrease
in $g_{2}$ at $r=2D^{+}$. Although the first discontinuity is less
pronounced and statistics are not good enough to unambiguously determine
its shape, it appears that it also has the same shape as the second
discontinuity, only of smaller magnitude. The split second peak is
of great importance because it is a clear signature of the strong
local order in the first two coordination shells of the packing, and
in fact observations have been made that along with the appearance
of a peak in $P_{f}(f)$ for small forces, the splitting of the second
peak of $g_{2}$ is a signature of jamming \cite{ForceChains_OHern}.
It is therefore important to try to understand the local geometrical
patterns responsible for the occurrence of these structures in $g_{2}$.

\subsubsection{Contact-Network Statistics}

\begin{figure}
\begin{center}\includegraphics[%
  width=0.90\columnwidth,
  keepaspectratio]{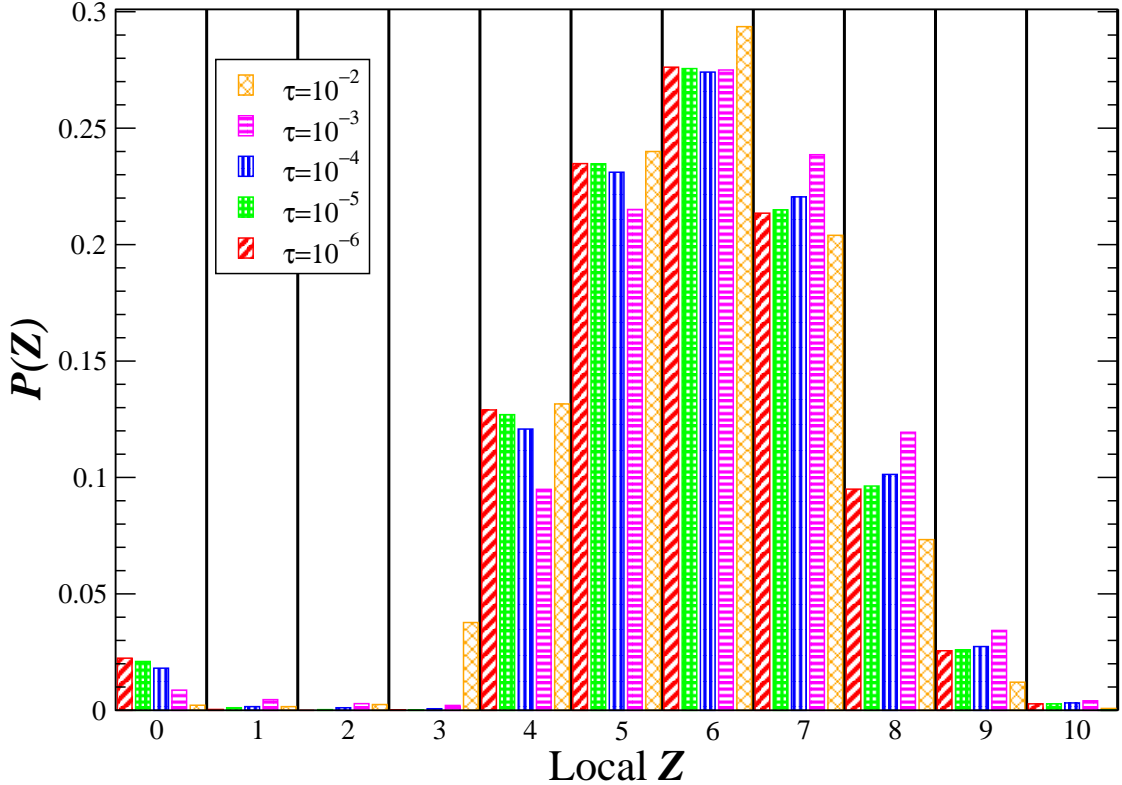}\end{center}

\caption{\label{P_Z.HS.N=3D10000}The probability distribution of local contact
numbers as the cutoff used in defining neighbors is increased. Rattlers
are clearly seen, and a relative maximum at $Z=6$ is seen. Note that
only one particle with 11 neighbors is observed, and very few have
as many as $10$ neighbors. No particle with 12 contacting neighbors
has been observed in any of our packings, indicating a lack of crystallinity.}
\end{figure}

The exact geometry of the jammed configuration $\V{R}_{J}$ is determined
(not necessarily uniquely) from its contact network, which as we have
demonstrated is the network of first-neighbor interactions and can
easily be separated from further-neighbor interactions. Fig. \ref{P_Z.HS.N=3D10000}
shows the histogram of local coordination numbers as a function of
the first-neighbor cutoff $\tau$, i.e., the histogram of the number
of particles within distance $(1+\tau)D$ from a given particle. It
is seen that for sufficiently small $\tau$ ($\tau<10^{-5}$) the
histograms are independent of the exact cutoff used (this is true
down to $\tau\approx100\delta$ or so, which can be as small as $10^{-12}$
in some of our packings). A number of particles having less than 2
contacts are seen, and these are clearly \emph{rattlers} and we have
removed them from consideration from all of the final packings we
analyze here. We observe that such particles remain with fewer than
2 contacts for a very wide range of $\tau$ and are easy to identify.
In some cases, however, we cannot unambiguously identify a handful
of the particles as rattlers or non-rattlers. This is typical for
packings which are not sufficiently close to their jamming point,
packings which have been produced using fast expansion in the LS algorithm,
or packings which are very large. It is safest to not remove such
particles as rattlers.

This work is the first time a clear look has been provided at the
exact contact network of disordered hard-sphere packings. Previous
studies have either used soft atoms, in which case the definition
of a contact is not clear cut unless one carefully takes limits of
a stiff interaction potential \cite{Friction_ForceChains}, and therefore
in such studies $\tau$ has been typically set to correspond to the
location of the first minimum in $g_{2}(r)$, or have used Voronoi
tessellations to define neighbors. Even studies which have actually
used hard particles have resorted to such definitions unsuitable to
investigating the jamming limit, mostly because the numerical precision
required to separate the true contacts from the near contacts has
not been achieved up to now \cite{IcosahedralOrder_Densification}.
Such investigations, the literature of which is too vast to cite,
have found a plethora of local coordination patterns typical of \emph{polytetrahedral}
packing, including icosahedral order \cite{IcosahedralOrder_Densification}.

We therefore attempted to do a similar \emph{shared-neighbor} \cite{IcosahedralOrder_Densification}
analysis for the contact networks of our disordered packings, and
look for local clusters reminiscent of polytetrahedral packing. Our
procedure, based on looking at the contact network as an undirected
graph, was as follows. For each particle, we extracted the subgraph
corresponding to the first neighbor shell of the particle (this includes
contacts between the neighbors), extracted its connected components,
and counted the number of occurrences of a given subgraph (using graph
algorithms that can test for graph isomorphism to form equivalence
classes). The results were surprising. By far the most prominent patterns
were a central particle contacting a \emph{chain} of 1, 2, 3, 4 or
5 contacting particles. The chains were almost never closed, other
than for chains of length 3 (which together with the original particle
form a contacting tetrahedron), and this was itself rare. The probability
of finding a chain of length $n$ seems to decay exponentially, $P(n)\sim\exp(-1.2n)$.
This study found very few tetrahedra, and so polytetrahedral local
ordering is certainly not apparent in the contact networks. We also
performed the same analysis for a range of $\tau$'s, all the way
up to $\tau=0.1D$ (which raises the average coordination significantly
above $6$), but still found the open linear chains to be the dominant
pattern. We further attempted to include second neighbors in the analysis,
however including all second neighbors led to very large subgraphs
of a very broad variety, so classification was not possible. We further
restricted our attention only to second neighbors which are very close
to the given particle (within $0.1D$, for example), and this also
found very few tetrahedra.

One of our goals was to determine if certain simple local coordination
patterns are responsible for each of the three features of $g_{2}(r)$
we previously documented: the power-law divergence near contact, and
the discontinuous, if not diverging peaks at $r=\sqrt{3}D$ and $r=2D$.
We had little success in accounting for the first one by restricting
attention to only the first two neighbor shells in the true contact
network. In particular, we looked at all the near contacts (for example,
with gaps less than $0.01D$) and whether the almost contacting particles
were in fact second neighbors in the contact network. Indeed most
were, however the majority only shared one particle as a first neighbor,
or two or three first neighbors which were not themselves first neighbors.
It was therefore not possible to isolate one particular local geometry
as responsible for the multitude of near contacts. An interesting
quantity we measured is the probability distribution $P_{\theta}(\theta)$
of bond-pair angles $\theta$ in the contact network, meaning the
angles between two contact bonds of a given particle. This distribution
is shown in the inset in Fig. \ref{g2.HS.split-peak}, and shows divergences
at $\theta=\pi/3$ and $\theta=2\pi/3$, which correspond to distances
$r=2D\sin(\theta/2)$ of $r=D$ and $r=\sqrt{3}D$. Although there
is no divergence at $\theta=\pi$, the corresponding distribution
of distances does show a divergence at $r=2D$.

We had more success with a shared-neighbors analysis for the split
second peak. This was because we could increase $\tau$ and thus progressively
relax the definition of first neighbor. We found that with increasing
$\tau$, an increasing majority of particle pairs at a distance close
to $\sqrt{3}D$ were second neighbors, and that an increasing majority
of them shared two neighbors which were themselves neighbors. This
corresponds to two edge-sharing \emph{approximately} equilateral coplanar
triangles, a configuration which has been suggested as being responsible
for the first part of the split second peak \cite{RCP_SerialDeposition}.
Note however that we do not observe any discontinuity in $g_{2}$
at $r=1.633D$, which corresponds to two face-sharing tetrahedra,
which is another configuration often mentioned in connection with
the split second peak. A similar analysis for the peak at $2D$ indicated
that the majority of particle pairs at this distance share one neighbor
between them, which represents an approximately linear chain of three
particles, a configuration which has long been known to be responsible
for the second part of the split second peak of $g_{2}$.

\subsection{Ordered Packings}

In this work we have focused on disordered hard-sphere packings, and
have found a multitude of unexpected singular features, such as a
long power-law tail in $g_{2}^{(\delta)}(l)$, a nonzero $P_{f}(f=0)$,
and a power-law divergence in $g_{2}^{(b)}(l)$. It is important to
realize that the properties we observe are not universal and will
change as one changes the amount of ordering of the packings. In particular,
dense ordered packings like the FCC crystal are not isostatic, and
we have no theory that can predict the shape of $g_{2}^{(\delta)}$.
We therefore resort to a computational investigation of ordered and
partially ordered sphere packings.

\subsubsection{Vacancy-Diluted FCC Crystal Packings}

\begin{figure}
\begin{center}\includegraphics[%
  width=0.90\columnwidth]{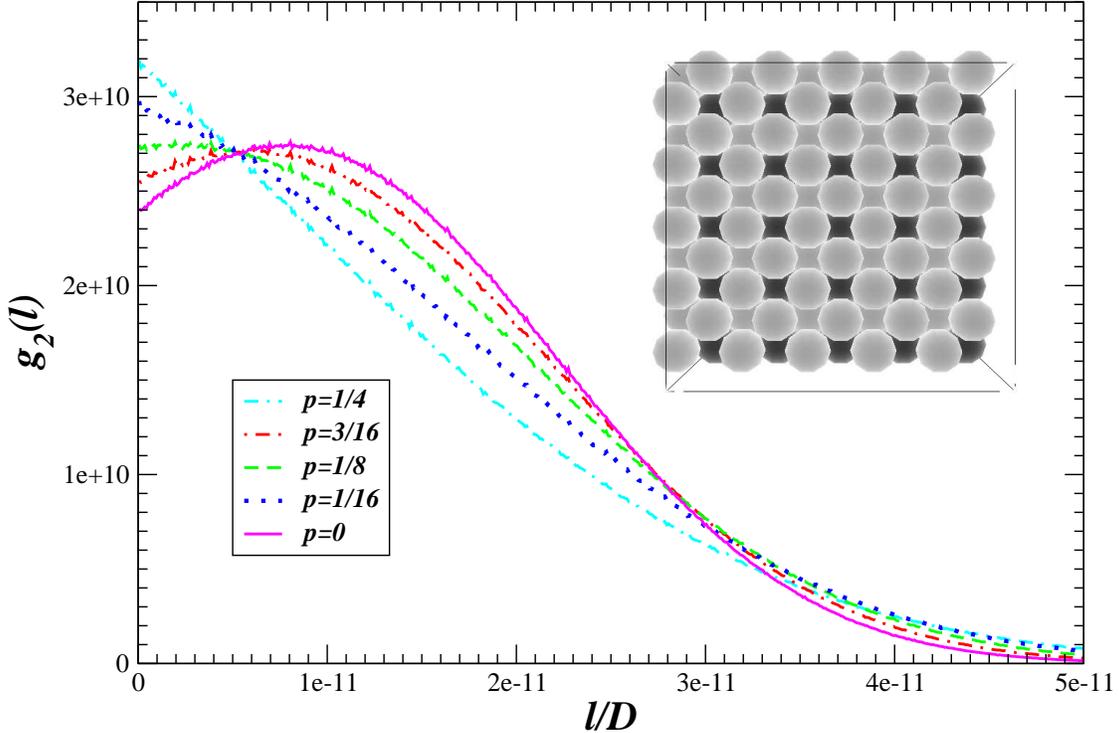}\end{center}

\caption{\label{g2.HS.FCC}The first-shell $g_{2}^{(\delta)}(l)$ for a collection
of FCC crystal packings with a fraction $p$ of the spheres removed,
starting with $N=13,500$ particles. The inset shows the packing with
most vacancies, where every $4^{\textrm{th}}$ sphere is removed to
form a cubic sublattice of vacancies (colored dark). Intermediate
$p$'s are achieved by randomly adding back some of the spheres to
the sublattice. The density has been reduced by $\delta=\sqrt{2}\cdot10^{-11}$
from close packing.}
\end{figure}

It was the behavior of crystal packings around the jamming point that
was the subject of Refs. \cite{FreeVolume_ClosePacked} and \cite{LimitingPolytope_HS},
and these works inspired this investigation. For crystal packings,
there is no ambiguity in defining first neighbors, and the FCC packing
has $Z=12$ contacts per particle, which is twice the isostatic value.
Therefore, the limiting polytope $\mathcal{P}_{\V{x}}$ is not a simplex
and, as argued in Ref. \cite{LimitingPolytope_HS}, it is expected
that for an FCC packing $g_{2}^{(\delta)}(l)$ will have a single
peak for small gaps. We indeed observe this computationally, for the
first time, as shown in Fig. \ref{g2.HS.FCC}. 

Furthermore, we have prepared vacancy-diluted FCC packings by removing
a fraction $p$ of the spheres from a perfect crystal, $0\leq p\leq4$
(here $p=0$ corresponds to the perfect crystal). The FCC lattice
is composed of 4 interpenetrating cubic lattices. We obtain the vacancy-diluted
crystal with the lowest density by removing one of these 4 cubic lattices
(i.e., $p=1/4$), as shown in the inset in Fig. \ref{g2.HS.FCC}.
This gives a packing with density of about $\phi\approx0.56$ and
mean coordination $\bar{Z}=8$ and is still collectively jammed. In
fact, it is likely that more spheres can be removed with a more elaborate
procedure \cite{Anu_order_metrics}. We can add back a randomly chosen
fraction $q=1/4-p$ of the previously removed quarter of the spheres,
to obtain $0<p<1/4$. The delta-function contributions to $g_{2}$
for several $p$'s are shown in Fig. \ref{g2.HS.FCC}. It is rather
surprising that the pair correlation function for the $p=1/4$ packing
no longer shows a peak, but is monotonically decaying. In fact, by
changing $p$ one can obtain packings with $g_{2}^{(\delta)}(l)$
that has zero slope at the origin. 

\begin{figure}
\begin{center}\includegraphics[%
  width=0.75\columnwidth,
  keepaspectratio]{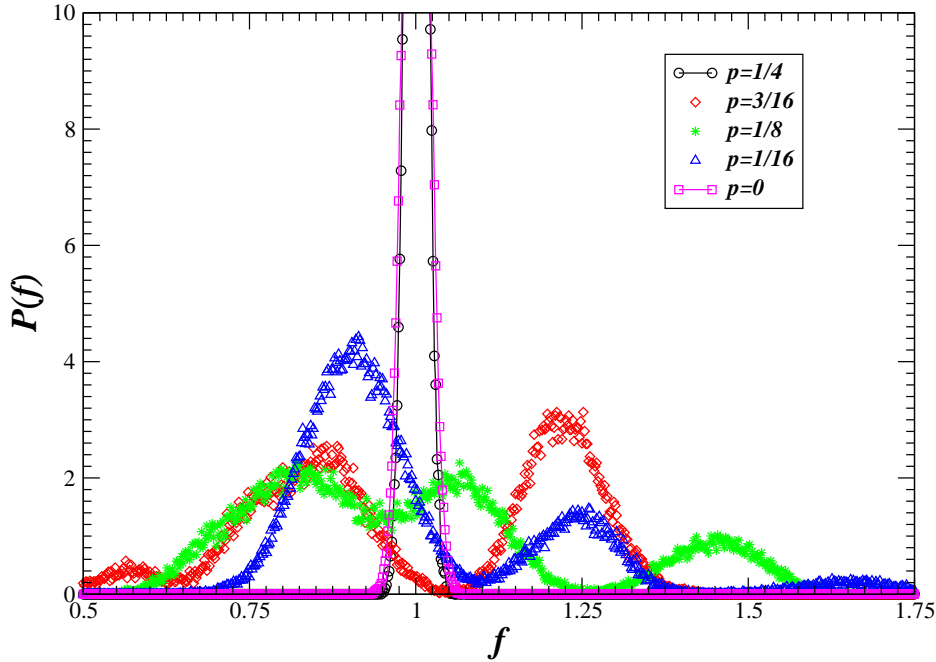}\end{center}

\caption{\label{P_f.FCC.sublattice}The force probability distribution for
the collection of FCC crystal packings from Fig. \ref{g2.HS.FCC}.
For the pure crystal and the crystal with most vacancies, all of the
particle pairs are identical and therefore the probability distribution
would be a delta function if forces are averaged over an infinite
time horizon. For the intermediate $p$'s, multiple relatively broad
peaks are observed. In contrast with the disordered case, very small
forces are \emph{not} observed.}
\end{figure}

It is interesting to note that for the (vacancy-diluted) FCC packings
$g_{2}^{(\delta)}(l)$ decays in a Gaussian manner, and in fact is
perfectly fitted by a modified Gaussian, $g_{2}^{(\delta)}(l)=(Al^{2}+Bl+C)\exp\left[(l-D)^{2}\right]$.
This fast decay is to be compared to the slow power-law decay for
the disordered packings {[}c.f. Eq.(\ref{Laplace_transform}){]},
hinting at possible connection to the stability of the crystal packings
versus the metastability of the glass packings \cite{GlassTransition_Rintoul}.
Additionally, we show the force distribution $P_{f}(f)$ for these
ordered packings in Fig. \ref{P_f.FCC.sublattice}, illustrating that,
in contrast with the disordered packings, very small forces are not
observed. It would be interesting to know if the perfect FCC crystal
can be vacancy-diluted to an isostatic packing and still be collectivelly
or strictly jammed, and what the corresponding force distribution
would be.

\subsection{Partially Crystallized Packings}

\begin{figure}
\begin{center}\includegraphics[%
  width=0.75\columnwidth,
  keepaspectratio]{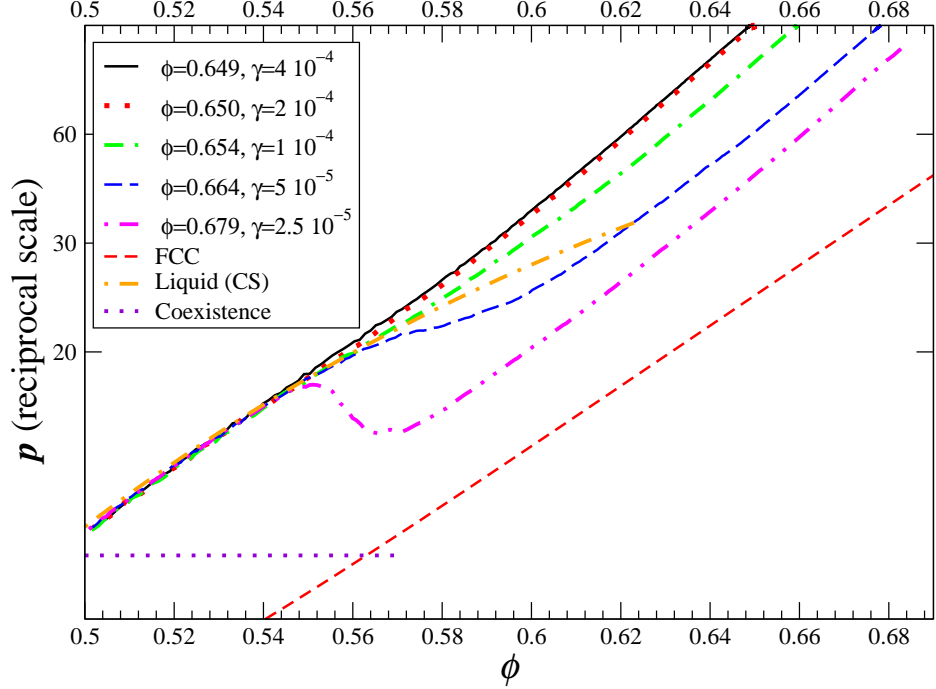}\end{center}

\caption{\label{PV.HS.N=3D10000.compression.from=3D0.5}Compression of an
initially liquid system with $\phi=0.5$ to jamming with several different
expansion rates $\gamma$ (the mean thermal velocity is 1, in comparison).
The pressure is plotted on a reciprocal scale (the tickmarks being
equally spaced in equal increments of $p^{-1}$, increasing in the
usual direction), to highlight the expected linear relation (\ref{PV_free_volume})
near jamming. The pressure-density curves for the perfect FCC crystal
\cite{Solid_Branch_HS}, the accepted fluid-solid coexistence region,
and the widely known Carnahan-Starling equation of state for the fluid
branch, are also shown for comparison. Sufficiently fast compression
suppresses crystallization and leads to densities around $0.64-0.65$,
and slower compression allows for partial crystallization, typically
occurring around $\phi\approx0.55$, which is the end of the coexistence
region (i.e., the density where the crystal necessarily becomes thermodynamically
favored). This produces denser packings which exhibit more crystal
ordering the denser they are. }
\end{figure}

\begin{figure}
\begin{center}\includegraphics[%
  width=0.75\columnwidth,
  keepaspectratio]{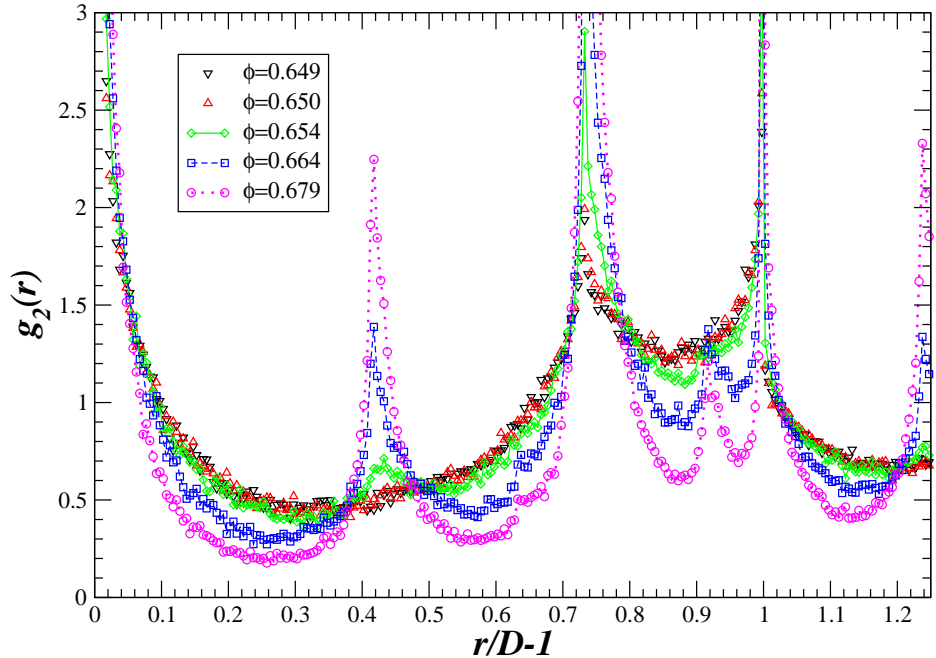}\end{center}

\caption{\label{g2.HS.ordered.split-peak}The evolution of the peaks in $g_{2}(r)$
as crystalline order is increased, for the packings from Fig. \ref{PV.HS.N=3D10000.compression.from=3D0.5}.
The formation of peaks at distances typical of the FCC lattice, such
as $r=\sqrt{2}$, is clearly seen. It is interesting to note that
a peak is observed at $\sqrt{11/3}\approx1.91$, which is a fifth-neighbor
distance in the HCP (but not the FCC) lattice (a similar HCP peak
at $\sqrt{8/3}\approx1.63$ is barely visible)}
\end{figure}

As previously explained, the Lubachevsky-Stillinger algorithm can
produce partially crystalline sphere packings if a sufficiently small
expansion rate is used and nucleation of crystallites occurs during
the densification. This is demonstrated in Fig. \ref{PV.HS.N=3D10000.compression.from=3D0.5},
where we show the evolution of the pressure during the densification
of an initially liquid sample (i.e., a state on the stable equilibrium
liquid branch), for a range of expansion rates $\gamma$. The slower
the expansion is, the more crystalline the final packings become,
as can be seen from the fact that the final density increases and
from the evolution of the peaks in $g_{2}(r)$, as shown in Fig. \ref{g2.HS.ordered.split-peak}.
Additionally, the structure factor $S(\V{k})$ shows more anisotropy
and localized peaks.

\begin{figure}
\begin{center}\includegraphics[%
  width=0.75\columnwidth,
  keepaspectratio]{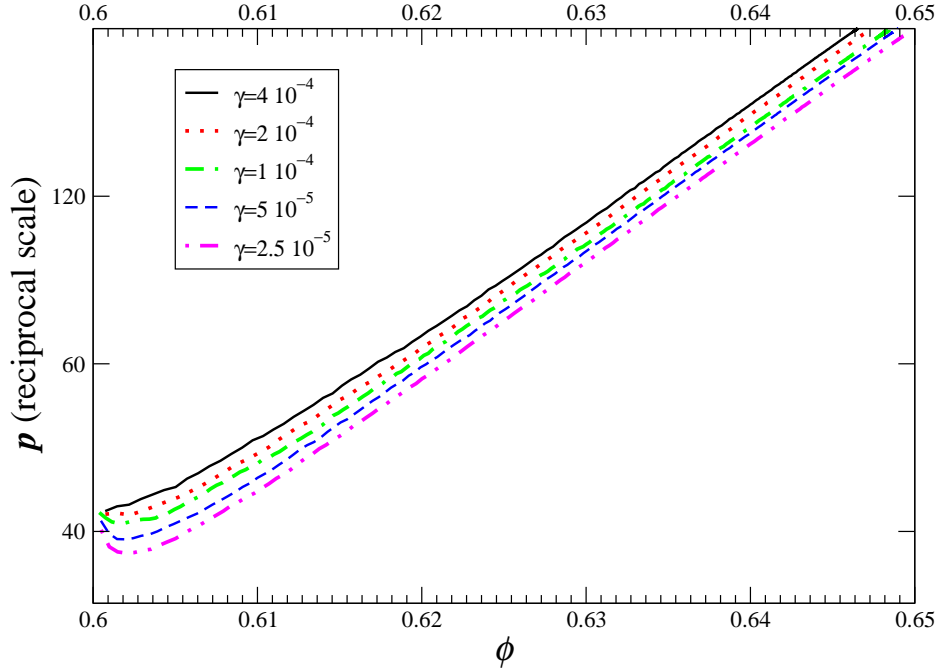}\end{center}

\caption{\label{PV.HS.N=3D10000.compression.from=3D0.6}Compression of an
initially (metastable) liquid system with $\phi=0.6$ to jamming with
several different expansion rates, as in Fig. \ref{PV.HS.N=3D10000.compression.from=3D0.5}.
For this range of expansion rates, crystallization is suppressed due
to the large initial density and all final packings are apparently
disordered and would be ordinarily identified as random, however,
it is clear that slower compression leads to higher densities, and
thus the final packings are not all identical, but rather, some are
more ordered than others, as can be verified by the slight increase
in bond-orientational order metric $Q_{6}$ \cite{Torquato_MRJ},
for example.}
\end{figure}

The packings shown in Fig. \ref{PV.HS.N=3D10000.compression.from=3D0.5}
clearly have nucleated crystals, and so one may anticipate that there
is a qualitative distinction between them and the {}``random'' packings
produced by suppressing crystallization. However, as demonstrated
in Fig. \ref{PV.HS.N=3D10000.compression.from=3D0.6}, slower densification
leads to larger densities and more ordered packings even if crystallization
is suppressed and no visible nucleation occurs. This indicates that
there is a continuum of packings from most disordered to perfectly
ordered \cite{Torquato_MRJ} packings, so that one needs to be careful
in interpreting results obtained from packings produced by just one,
possibly non-trivially biased, algorithm. For example, Ref. \cite{ForceChains_OHern}
relates the occurrence of a peak in $P_{f}(f)$ to jamming. However,
as we show next, jammed packings do not necessarily exhibit this peak
if they are sufficiently ordered.

For the sake of brevity, we will only briefly discuss some interesting
features of $g_{2}$ for the partially crystallized packings. Since
the perfect FCC/HCP crystals have $\bar{Z}=12$, one expects that,
as partial crystallization occurs, somehow the number of first neighbors
per particle should increase from the isostatic value of $\bar{Z}=6$.
However, this is not really so if one \emph{properly} defines first
neighbors via true contacts in the final jammed packing. In fact,
if one plots $Z(l)$ for partially crystallized packings (we omit
this plot), a qualitatively similar curve to that shown in Fig. \ref{G2.HS.compression}
is seen, with $\bar{Z}$ clearly close to the isostatic value of $6$.
However, the background $Z^{(b)}(l)$ shows a faster rise the more
crystalline the packing is {[}consistent with a larger exponent $\alpha$
as defined in Eq. (\ref{Z_power_law}){]}, so that indeed an increase
of the cumulative coordination is seen for sufficiently large gaps.
Additionally, we observe that nearly crystalline packings easily jam
in noticeably hypostatic configurations, with a higher probability
of observing particles with only $2$ or $3$ contacts and a less
flat plateau in $Z(l)$.

\begin{figure}
\begin{center}\includegraphics[%
  width=0.75\columnwidth,
  keepaspectratio]{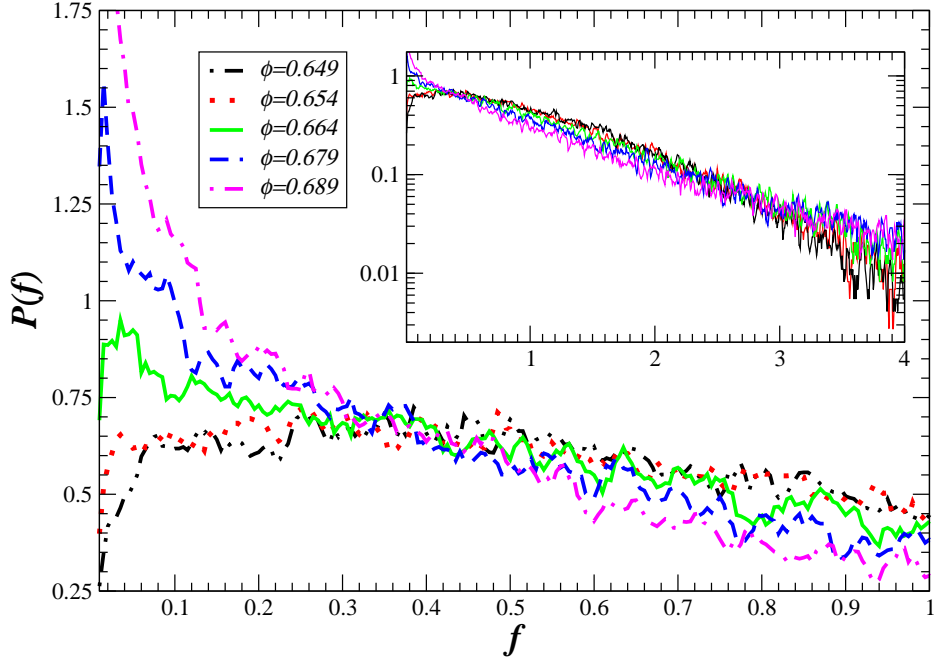}\end{center}

\caption{\label{P_f.HS.crystalline}The evolution of $P_{f}(f)$ as crystalline
order is increased, showing the disappearance of the peak at small
forces. The inset shows a log-log view of the plot, and is consistent
with exponential decay for large forces.}
\end{figure}

All of these findings are readily explained. The basic premise, used
widely in the granular media literature, is that random perturbations
to either the particle-size distribution or to the boundary conditions
will break some of the contacts in an otherwise perfect crystal down
to the isostatic value. This is because additional contacts in excess
of $\bar{Z}=6$ imply special correlations between the positions of
the particles, which one expects to destroy with random perturbations.
Such random perturbations are provided in the case of partially crystallized
packings by the fact that the crystallites need to jam against a partially
amorphous surroundings, and this induces complex strains that break
some of the perfect-crystal contacts.%
\footnote{We mention in passing that we have observed similar results by starting
with a perfect FCC crystal, applying a small (but not too small) random
strain, and then jamming the packings. This typically yields almost
perfectly crystal packings which are nonetheless clearly frustrated
by the random strain to have $\bar{Z}\approx6$.%
} However, the geometric peculiarities of the underlying crystal remain;
for example, there is a multitude of nearly collinear (in fact lines
of aligned particles) or coplanar contacts, which leads to the occurrence
of much more pronounced \emph{force chains} (chains of large forces
propagating along a nearly straight line) and a sharp increase in
the probability of occurrence of small forces. We indeed observe this
in Fig. \ref{P_f.HS.crystalline}, where we show that for sufficiently
ordered packings there is no longer a peak in $P_{f}(f)$ for small
forces, but rather a monotonic decrease of $P_{f}(f)$, apparently
exponential for sufficiently large forces. This is in contrast to
previous studies of the effect of order on force distributions in
granular piles \cite{ForceChains_Ordering,ForceChains_Walls}, which
did not register a significant impact of the ordering. However, these
studies study the distribution of forces in granular piles and a direct
comparison is beyond the scope of this work.

\section{Discussion}

The results presented in this work settle some long-standing questions
and confusions in the literature. For the first time, we showed both
theoretically and computationally how the delta-function portion of
$g_{2}(r)$ is formed as jamming is approached, for a true hard-sphere
packing. Our investigation focused on maximally disordered (MRJ) sphere
packings with a packing fraction $\phi\approx0.64-0.65$. We presented
the first true hard-sphere computational data on the power-law divergence
in the near-contact portion of $g_{2}$, in agreement with previous
observations in the literature for stiff soft spheres, but with a
distinguishably different coefficient of $-0.4$. We confirmed that
this divergence persists even in the true jamming limit for hard particles.
We presented high-quality data on the probability distribution of
interparticle forces $P_{f}(f)$, especially focusing on small forces,
demonstrating a maximum at small forces and a nonzero intercept at
$f=0$. A local analysis of the topology of the contact network found
few traces of tetrahedra and an overwhelmingly complex local connectivity,
and was successful in accounting for the structures responsible for
the split second peak of $g_{2}(r)$. A computational study of the
delta-function contribution to $g_{2}(r)$ for vacancy-diluted FCC
crystals showed a faster than exponential decay, unlike the slow power-law
decay for the disordered isostatic packings. Finally, we investigated
packings on the transition from maximally disordered to maximally
ordered, and found that partially crystallized packings produced by
our algorithm are still nearly isostatic despite having a higher density,
and that $P_{f}(f)$ loses the peak for sufficiently ordered packings.

This work has raised several important questions. The computational
observations undermine the very applicability of the ideal jammed
packing model to large (maximally) disordered packings of spheres,
as produced by most algorithms in use today. First, a very unusual
power law divergence in $g_{2}(l)$ is observed near contact, leading
to a multitude of particle pairs just away from contact. Similarly,
a power-law decay is seen in the contact part of $g_{2}(l)$. As the
packings become larger, one can expect the tails of the two power
laws to start overlapping by an observable number of contacts, blurring
the distinction between true contacts and almost contacts. Even more
troubling is the observation that there appears to be a positive probability
of observing a zero force in the contact network of the packings,
indicating the presence of geometric degeneracies in the contact network.
The above observations may explain why we have had trouble generating
truly jammed packings of $N=10,000$ particles. However, we do not
see a reason why very large but finite packings collectively jammed
ideal packings could not be constructed. The question of what algorithm
can produce disordered (and thus likely isostatic) packings which
are jammed and devoid of some or all of the above peculiarities, as
is the FCC crystal packing%
\footnote{Note that the observations we list as troubling are separate from
the rather general objections due to the inapplicability of the concept
of ideal jamming to infinite packings, which apply to crystal packings
as well \cite{FreeVolume_ClosePacked}.%
}, for example, remains open. As usual, with each careful study the
hard-sphere system provides more questions than originally posed or
answered!

\end{document}